\documentclass[aps,prb,twocolumn,superscriptaddress]{revtex4-2}
\usepackage{graphicx}
\usepackage{float}
\usepackage{multirow}
\usepackage{hhline}
\usepackage[table]{xcolor}
\usepackage{amsmath}
\usepackage{amssymb}
\usepackage{dcolumn}
\usepackage{bm}
\begin{document}
	
	\title{First-principles calculations of double resonance Raman spectra for monolayer MoTe$_2$}
	
	\author{Jianqi Huang}
	\affiliation{Shenyang National Laboratory for Materials Science, Institute of Metal Research, Chinese Academy of Sciences, School of Material Science and Engineering, University of Science and Technology of China, Shenyang 110016, P. R. China}
	\author{Huaihong Guo}
	\email{hhguo@alum.imr.ac.cn}
	\affiliation{College of Sciences, Liaoning Petrochemical University, Fushun 113001, P. R. China}
	\author{Lin Zhou}
    \affiliation{School of Chemistry and Chemical Engineering, Frontiers Science Centre for Transformative Molecules, Shanghai Jiao Tong University, Shanghai 200240, P. R. China}
    \author{Shishu Zhang}
    \affiliation{Center for Nanochemistry, Beijing Science and Engineering Center for Nanocarbons, Beijing National Laboratory for Molecular Sciences, College of Chemistry and Molecular Engineering, Peking University, 100871, Beijing, P. R. China}
    \author{Lianming Tong}
    \affiliation{Center for Nanochemistry, Beijing Science and Engineering Center for Nanocarbons, Beijing National Laboratory for Molecular Sciences, College of Chemistry and Molecular Engineering, Peking University, 100871, Beijing, P. R. China}
    \author{Riichiro Saito}
    \affiliation{Department of Physics, Tohoku University, Sendai, Miyagi 980-8578, Japan}
	\author{Teng Yang}
	\email{yangteng@imr.ac.cn}
	\affiliation{Shenyang National Laboratory for Materials Science, Institute of Metal Research, Chinese Academy of Sciences, School of Material Science and Engineering, University of Science and Technology of China, Shenyang 110016, P. R. China}
	\author{Zhidong Zhang}
	\affiliation{Shenyang National Laboratory for Materials Science, Institute of Metal Research, Chinese Academy of Sciences, School of Material Science and Engineering, University of Science and Technology of China, Shenyang 110016, P. R. China}
	
	\begin{abstract}
		Since double resonance Raman (DRR) spectra are laser-energy dependent, the first-principles calculations of DRR for two-dimensional materials are challenging. Here, the DRR spectrum of monolayer
		MoTe$_2$ is calculated by home-made program, in which we combine {\em ab-initio} density-functional-theory calculations with the electron-phonon Wannier (EPW) method. Within the fourth-order perturbation theory, we are able to quantify not only the electron-photon matrix elements within the dipole approximation, but also the electron-phonon matrix elements using the Wannier functions. The reasonable agreement between the calculated and experimental Raman spectra is achieved, in which we reproduce some distinctive features of transition metal dichalcogenides (TMDCs) from graphene (for example, the dominant intervalley process involving an electron or a hole). Furthermore, we perform an analysis of the possible DRR modes over the Brillouin zone, highlighting the role of low-symmetry points. Raman tensors for some DRR modes are given by first principles calculations from which laser polarization dependence is obtained.
	\end{abstract}
	
	\maketitle
	
	\section{introduction}
	Raman spectroscopy as a versatile probe tool has been widely used for characterizing a broad range of physical properties including superconductive~\cite{Measson2014}, topological~\cite{Kung2017}, ferroelectric~\cite{Tenne2006} properties, magnetic ordering~\cite{Huang2020} and phase transition~\cite{Kim2019}, electronic interference effect~\cite{Miranda2017,Zhang2022}, phonon helicity~\cite{Tatsumi2017}. With the rise of two-dimensional (2D) van der Waals materials, Raman spectroscopy plays an essential role in supplying information on the heterostructure and intrinsically topological properties~\cite{Chen2016,Zhang2016,Liu2018,Zhang2020}. In particular, the second-order Raman spectra, which have been widely observed in transition metal dichalcogenides (TMDCs) and other semiconducting 2D materials ~\cite{Wakabayashi75,Sourisseau1991,Sekine84,Stacy85,Chen74,Sourisseau89,Sourisseau1991,Feldman96,Frey99,Windom11,Li12,Chakraborty13,Terrones14,Berkdemir13,Guo15,Livneh2015,Carvalho2017}, host non zone-centered two phonons, which are strongly dependent on the laser excitation energy. Double resonance Raman (DRR) process is essential for observing the two-phonon Raman spectra or defect-oriented Raman spectra, whose intensity is comparable to or even larger than the resonant Raman process of a zone-centered phonon~\cite{Ferrari2013,Liu2015}.
	
    However, the first principles calculation of the DRR spectra has been a long-term challenge before it is possible to accurately quantify electron-phonon matrices with sufficiently dense grid under Wannier interpolation throughout the Brillouin zone (BZ)~\cite{Giannozzi2017}. With the developed electron-phonon Wannier technique, Herziger \textit{et al.}~\cite{Herziger2014} and Torche \textit{et al.}~\cite{Torche2017} have calculated the double-resonant 2D mode of graphite by first-principles calculations. However, compared to the 2D overtone band in graphite which is intensively studied and has prior assignment due to empirical method, it remains challenging to identify by first principles calculation the spectra of a system without prior double-resonance information. It is still necessary to calculate the DRR scattering amplitudes for all electron wavenumbers, $k$, for all phonon wavevectors, $q$ with the given $k$, and for all combinations of two phonon modes and electronic energy subbands as a function of laser excitation energies.

    So far, the DRR analysis in TMDC is largely semi-quantitative or even qualitative, due to the lack of a quantitative treatment of electron-phonon coupling. The earliest interpretations of DRR of TMDCs were merely by comparing phonon frequencies with reference to some inelastic neutron scattering results~\cite{Wakabayashi75,Sourisseau1991}, which therefore leads to some inconsistency in mode assignments~\cite{Sekine84,Stacy85,Chen74,Sourisseau89,Sourisseau1991,Feldman96,Frey99,Windom11,Li12,Chakraborty13}. Terrones \textit{et al.}~\cite{Terrones14} and Berkdemir \textit{et al.}~\cite{Berkdemir13} assigned phonon modes in few-layer WSe$_2$ by referring to both the band structure and the high-symmetry point phonon. Guo \textit{et al.}~\cite{Guo15} went further to specify the phonon wave vectors around the high-symmetry M point for the DRR electron-phonon resonance in MoTe$_2$. Carvalho \textit{et al.}~\cite{Carvalho2017} measured the second-order bands using more than twenty different laser excitations in conjunction with DFT calculations. Livneh \textit{et al.}~\cite{Livneh2015} made a comprehensive multiphonon spectral analysis in MoS$_2$ based on group theory. All the recent endeavors have been either focused on the high-symmetry points in the reciprocal space, or treating the electron-phonon matrix element as constant, which can hardly reveal the essential message on the interplay between electron and phonon. Tatsumi \textit{et al.}~\cite{Tatsumi2017} made a computer program of calculating first-order resonance Raman spectra by first principles calculation in which electron-photon and electron-phonon matrix elements are calculated by the wave function coefficient and Electron-Phonon-Wannier (EPW) for zone-center phonon mode. Their concept can be extended for DDR spectra, though a huge amount of calculations is required.
	
	In this paper, we study theoretically the second-order Raman process of monolayer MoTe$_2$. We developed a computer program for calculating the DDR spectra by first principles calculation in which we adopt Quantum-Espresso (QE) and EPW for evaluating the electron-photon and electron-phonon matrix elements, respectively, and compare with the experimental Raman spectra for several excitation energies. The reasonable agreement between the calculated and experimental double resonance Raman spectra is achieved, showing the reliability of our method and revealing some distinctive features of TMDCs from graphene. Furthermore, assignment of DRR modes on any possible combinations of two-phonon modes at any random $k$ point is directly obtained. Raman tensors for some DRR modes are obtained from which we can discuss polarization dependence of two-phonon Raman spectra.
	
	\section{methods}
	
	\subsection{computational methods}
	Second-order double resonance Raman intensity with Raman shift $E_{RS}$ as a function of the incident laser energy $E_L$ can be described as following expression,
	\begin{equation}\label{eq1}
		\begin{split}
			I\left(E_L\right)\propto&\sum_{\textbf{q},\mu,\nu}\left|\bm{P}_s^{\dagger}\cdot\stackrel{\leftrightarrow}{R}\left(\bm{q},\mu,\nu, E_L\right)\cdot\bm{P}_i\right|^2 \\ &\delta\left(E_{RS}\pm\hbar\omega_\mu\pm\hbar\omega_\nu\right),
		\end{split}
	\end{equation}
	where $\bm{P}_i$, $\bm{P}_s$ represent the polarization directions of incident light and scattered light, respectively, $\omega_\mu$ and $\omega_\nu$ denote the phonon frequencies corresponding to two phonon modes $\mu$ and $\nu$. The Raman tensor takes a fourth-order perturbation form of
	\begin{widetext}
		\begin{equation}
			\stackrel{\leftrightarrow}{R}\left(\bm{q},\mu,\nu, E_L\right) = \sum_{\bm{k},i=f,n,n',n''}
			\frac{\bm{D}_{fn''}\left(\bm{k}\right)\cdot M_{n''n'}^{ep,\nu_{-\bm{q}}}\left(\bm{k}+\bm{q}\right)\cdot M_{n'n}^{ep,\mu_{\bm{q}}}\left(\bm{k}\right)\cdot\bm{D}_{ni}^{\dagger}\left(\bm{k}\right)}{\left(E_{ni}-E_L-\mathrm{i}\gamma\right)\left(E_{n'i}-E_L\pm\hbar\omega_\mu-\mathrm{i}\gamma\right)\left(E_{n''i}-E_L\pm\hbar\omega_\mu\pm\hbar\omega_\nu-\mathrm{i}\gamma\right)},
			\label{eq2}
		\end{equation}
	\end{widetext}
	where $\bm{D}_{fn''}$ is the electric dipole vector $\langle f| \bm{D} | n''\rangle$ and $M_{n''n'}^{ep,\nu_{-\bm{q}}}$ is the electron-phonon coupling matrix elements. $i$, $n$, $n'$, $n''$ and $f$ denote, respectively, the initial state, the three intermediate states, and the final state of an electron. Since the broadening due to electron-phonon coupling is around 100 meV at room temperature, we adopt the value of damping constant $\gamma$ in our simulation. Since the backscattering configuration is set up in the experiment, the Raman spectra are calculated in the backscattering configuration ($\bar{\textrm{Z}}$(XX)Z).
	
	After obtaining the Raman spectrum, we take the following steps to assign the combined modes that contribute to a specific Raman peak. With fixing the Raman shift $E_{RS}$, we can obtain Raman intensity ($I_{\bm{q}}$) as a function of $q$ in the Brillouin zone,
	\begin{eqnarray}
		 I_{\bm{q}}&\propto&\sum_{\mu\nu}\left|\bm{P}_s^{\dagger}\cdot\stackrel{\leftrightarrow}{R}\left(\bm{q},\mu,\nu\right)\cdot\bm{P}_i\right|^2\delta\left(E_{RS}\pm\hbar\omega_\mu\pm\hbar\omega_\nu\right) \nonumber \\
		&\equiv& \sum_{\mu,\nu} I^{\mu\nu}_q .
		\label{eq3}
	\end{eqnarray}
	
	Meanwhile, with fixing two vibration modes $\mu$ and $\nu$ at the same time, we can obtain the Raman intensity ($I_{\bm{q}}^{\mu\nu}$) contributed by the combined two phonon modes as a function of $q$. Considering the entire Brillouin zone, the contribution of the fixed mode combination to total Raman intensity is achieved by summation
	\begin{equation}
		\zeta_{\mu\nu} =\sum_{\bm{q}}\frac{I_{\bm{q}}^{\mu\nu}}{I}.
		\label{eq4}
	\end{equation}
	Now, we can extract the ones that contribute significantly to the total Raman intensity by observing the value of $\zeta_{\mu\nu}$ of all possible mode combinations.
	
	We performed the electronic and phonon energy dispersion calculations on monolayer MoTe$_2$ by using first-principles density functional theory within the local density approximation (LDA) as implemented in the QE code~\cite{Giannozzi09}. The monolayer MoTe$_2$ are separated by 25 \AA\ from one another in a unit cell of the calculation to eliminate the inter-few-layer interaction. We used norm-conserving pseudopotentials (NCPP) within the local density approximation (LDA) with a plane-wave cutoff energy of 120 Ry to describe the interaction between electrons and ions. The spin-orbit split electronic band structures were calculated by the relativistic pseudopotentials derived from an atomic Dirac-like equation. The atomic coordinates were relaxed until the atomic force was less than 10$^{-5}$ Ry/Bohr. The Monkhorst-Pack scheme~\cite{Monkhorst76} was used to sample the Brillouin zone over a 16$\times$16$\times$1 and 8$\times$8$\times$1 $k$-mesh for electronic and phonon energy calculation, respectively. The phonon energy dispersion relations of MoTe$_2$ were calculated by the density functional perturbation theory~\cite{Baroni01}. On these basis, we calculated the electrical dipole vector by using a modified version of the QE code. Further by means of Wannier interpolation schemes as implemented in standard EPW~\cite{Noffsinger2010,Ponce2016}, we obtained the electron-phonon coupling matrix elements for each phonon mode on a much fine grid of 45$\times$45$\times$1 $k$-mesh in the Brillouin
	zone which is dense enough to achieve convincing results.
		
	\subsection{experimental method}
	Bulk crystals of MoTe$_2$ were prepared through a chemical vapor transport method~\cite{Lieth77}. Atomically thin crystals of MoTe$_2$ were mechanically exfoliated from the bulk crystals onto 90 nm SiO$_2$/Si substrate. Raman spectroscopy for monolayer MoTe$_2$ was performed using 532, 633 and 785 nm excitation lasers for discussing the observed phonon dispersion. The grating sizes were 1800 lines/mm for the 785, 633 and 532 nm laser excitation measurements. The magnification of the objective lens was 100x. The accumulation times were 60-300 seconds. All measurements were performed at room temperature in the backscattering configuration. Typical Raman spectra of monolayer 2H-MoTe$_2$ under different laser excitation are analyzed and compared with theoretical calculations in Fig.~\ref{Fig1} below.	
	
	\section{results and discussion}	
	\begin{figure*}[htbp]
		\includegraphics[width=1.95\columnwidth]{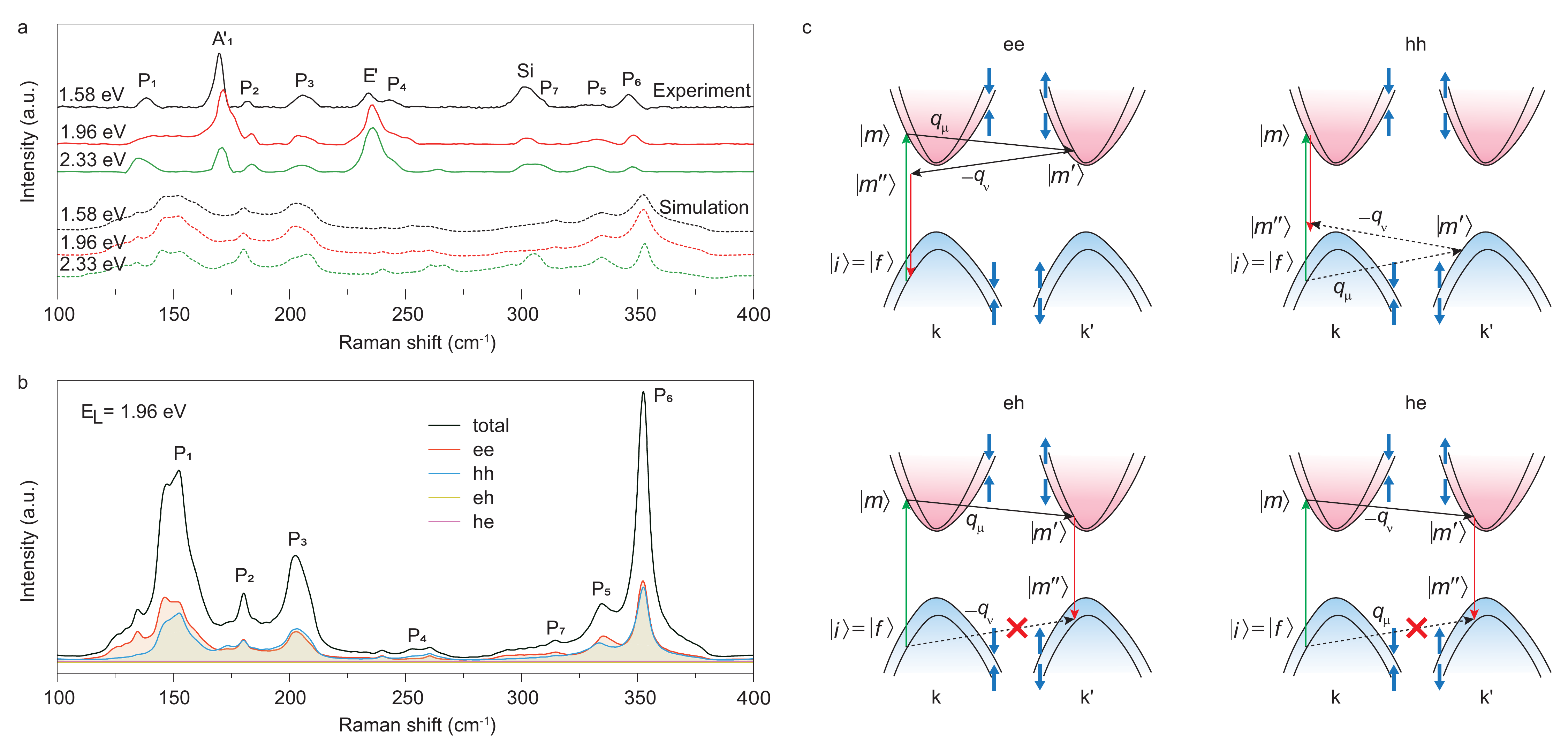}
		\caption{(Color online) Double Resonance Raman (DRR) process of monolayer MoTe$_2$. (a) Raman spectra of monolayer MoTe$_2$ under three different laser excitation energies, 2.33 eV (532 nm), 1.96 eV (633 nm) and 1.58 eV (785 nm), showing the small DRR intensity peaks P$_i$ (i = 1, 2, ..., 7). The experiment and simulated data are given in top and bottom panels by solid and dash lines, respectively. The typical first-order Raman peaks of A$^{\prime}_1$ (corresponding to A$_\textrm{1g}$ in bulk MoTe$_2$) and E$^{\prime}$ (E$_\textrm{2g}$) are only given in the experimental data. (b) Decomposition of total simulated Raman spectra into different Raman scattering process pathways, including the $ee$, $hh$, $eh$ and $he$ processes. (c) Schema of the four typical DRR pathways. $eh$ and $he$ are not allowed for the forbidden spin-flip in hole-phonon scattering.}
		\label{Fig1}
	\end{figure*}
	Fig.~\ref{Fig1}(a) shows Raman spectra of monolayer MoTe$_2$ under 2.33 eV (532 nm), 1.96 eV (633 nm) and 1.58 eV (785 nm) laser excitation energies from both experiment (top panel in solid lines) and simulation (bottom panel in dashed lines). The baseline correction of the raw experimental Raman spectra has been performed to remove the fluorescence noise. The experimental Raman spectra show two strong peaks, the in-plane E$^{\prime}$ mode at $\sim$ 236 cm$^{-1}$ (corresponding to E$_\textrm{2g}$ in bulk) and the out-of-plane A$^{\prime}_1$ mode at $\sim$ 171 cm$^{-1}$ (A$_\textrm{1g}$ in bulk) for monolayer MoTe$_2$. These spectra are assigned to the first-order Raman spectra of the $\Gamma$ point phonons~\cite{yamamoto14}. The calibration peak from Si at $\sim$ 300 cm$^{-1}$ is assigned to the 2TA mode~\cite{Spizzirri2010}. Since the present simulation method is only applicable for double resonance Raman peaks, the simulated Raman spectra can distinguish the DRR spectrum from the first-order peaks. As shown in Fig.~\ref{Fig1}(a), seven peaks are observed with relatively small intensities, which we denote as P$_i$ (i = 1, 2, ..., 7). The peak positions of each P$_i$ are found be dispersive as a function of excitation energy~\cite{Saito02}, either upshifting or downshifting by up to several cm$^{-1}$ by changing laser excitation energies, and were ascribed to the second-order Raman process~\cite{Guo15}. A reasonable agreement between experiment and our calculation substantiates the double resonance origin of these peaks, as evidenced in Fig.~\ref{Fig1}(a).
	
\begin{figure*}[htbp]
		\includegraphics[width=1.95\columnwidth]{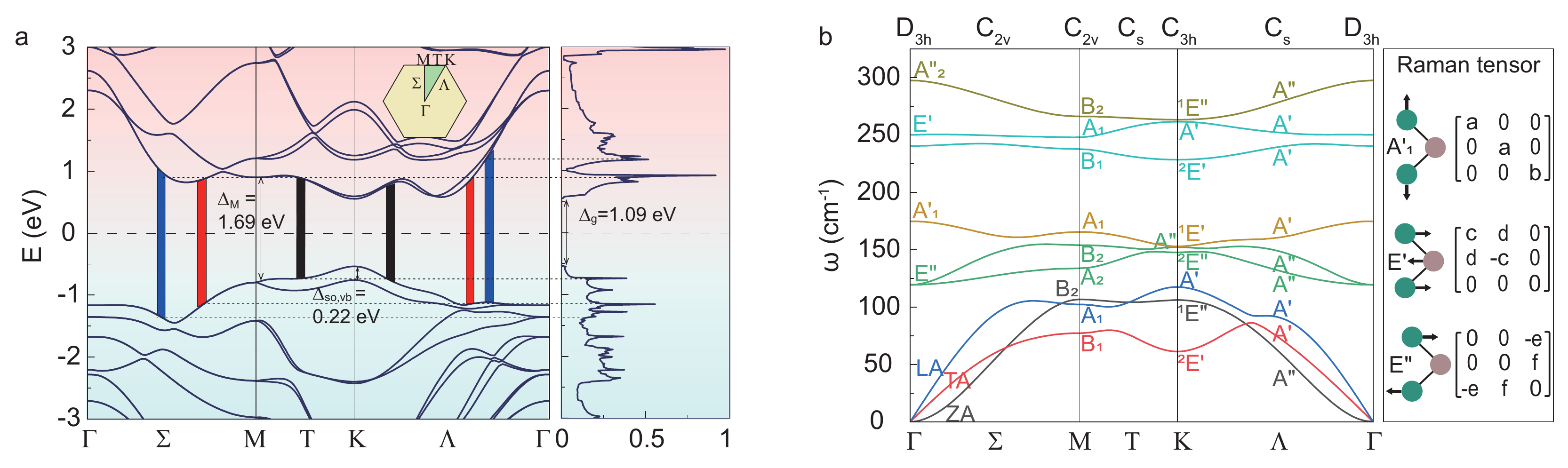}
		\caption{(Color online) (a) Electronic energy band structure and density of states (DOS) of MoTe$_2$. The vertical bar in black, red and blue represent the position of electronic photo-excitation due to laser excitation at energy of 1.58 eV, 1.96 eV and 2.33 eV, respectively. (b) The phonon dispersion relation of MoTe$_2$. The point groups are listed above for different $q$ points with the irreducible representations given on each colored band. Typical Raman-active $\Gamma$-point phonons are visualized with their Raman tensor given.}
		\label{Fig2}
	\end{figure*}

	Double resonance Raman process usually consists of several pathways~\cite{Venezuela11}. Depending on whether valence hole is involved and whether the optical absorption/emission occur at the same $k$ point, four typical pathways are $ee$ (only conduction electron involved), $hh$ (only valence hole involved), $eh$ and $he$ (electron and hole both involved), as indicated in Fig.~\ref{Fig1}(c). Our method can manage to decompose the contribution from each pathway to the Raman intensity quantitatively. In Fig.~\ref{Fig1}(b) taking the laser excitation 1.96 eV, for example, we compare the Raman spectra from each pathway with the total one. The $ee$ and $hh$ seem to contribute almost equally and dominantly to the total intensity, in contrast to the negligible contribution from the $eh$ and $he$ processes. Such a behavior is vastly different from what's observed in graphene~\cite{Venezuela11}, where $eh$ and $he$ dominate over $hh$ and $ee$. This is sensible considering that there doesn't exist electron-hole symmetry in MoTe$_2$, as indicated in the electronic band structure in Fig.~\ref{Fig2}(a), which strongly hinders both electron and hole scattering simultaneously by intervalley phonon. While in graphene with symmetric Dirac cone, intervalley scattering of both conduction electron and valence hole by phonon can take place in parallel~\cite{Venezuela11}. Further, the inversion-symmetry breaking and strong spin-orbit coupling (SOC) in monolayer MoTe$_2$ give rise to a large spin splitting on the electronic bands around the $K$ point, as shown in Fig.~\ref{Fig2}(a) and spin-valley locking occurs around the valley. Such special band features in MoTe$_2$ make the Raman transition of $eh$ and $he$ pathways unallowable because of the forbidding spin-flip in the hole-phonon scattering in such cases.
	
	In order to discuss the Raman active modes for two-phonon Raman scattering, we first make analysis on group theoretical selection rules for two-phonon scattering. Due to the dominant $ee$ and $hh$ processes for the DRR, the $\mu_{th}$ and $\nu_{th}$ two phonons contributing to a single DRR process should have the same wave vector with opposite sign to each other, q$_{\mu}$ and -q$_{\nu}$. The origin of DRR which can arise from either combination, or subtraction, or overtone of two phonon modes can therefore be analyzed at the same $q$ point in the phonon dispersion relation in Fig.~\ref{Fig2}(b). The point groups along high-symmetry line and $q$ points are put on top, such as point group D$_\textrm{3h}$ for the highest symmetric zone-center $\Gamma$ point, C$_\textrm{3h}$ for the zone corner $K$ point, C$_\textrm{2v}$ for the zone-edge middle $M$ point and between the $\Gamma$ and $M$ points, and C$_\textrm{s}$ for the rest of $q$ points (including between the $\Gamma$ (or $M$) and $K$ points) in the BZ. The irreducible representations are also given to the corresponding phonon bands. Three typical zone-center Raman-active modes (A$^{\prime}_1$, E$^{\prime}$ and E$^{\prime\prime}$) are visualized on the right of Fig.~\ref{Fig2}(b) with the Raman tensors. By the general methods of Birman~\cite{Birman1962,Birman1963}, with the help of the Bilbao crystallographic server~\cite{Aroyo2006}, we can determine whether two-phonon modes for a $q$ point are Raman-active or not, by correlating the irreducible representations of the combined modes of the group at the $q$ point with the irreducible representations of the full space group (D$_\textrm{3h}$) at the $\Gamma$ point. To ascertain the Raman activity of a two-phonon DRR mode, the reduced irreducible constituents from the Kronecker products have to contain at least one of the three Raman-active symmetries (A$^{\prime}_1$, E$^{\prime}$ and E$^{\prime\prime}$) of zone-center point group D$_\textrm{3h}$. In the Appendix Table~\ref{Table2}, we list all the possible combinations of two phonons in the BZ and the reduction of symmetries, and mark those Raman-active ones in blue color. First of all, the overtone modes, even from two Raman-inactive symmetries, are all Raman active in the whole BZ, since the decomposed symmetries include at least Raman-active A$^{\prime}_1$ symmetry. Second, for $q$ point beyond the zone-center $\Gamma$ point (at $M$ or between $\Gamma$ and $M$ of C$_\textrm{2v}$ group), especially with lower symmetry (C$_\textrm{s}$ group), the decomposed symmetries usually have two Raman-active symmetries (A$^{\prime}_1$ + E$^{\prime}$) coexisting. Considering that A$^{\prime}_1$ and E$^{\prime}$ have different circular polarization selectivity~\cite{Tatsumi2017}, we may anticipate that DRR modes at these points should have non-zero Raman intensity in both $\sigma_{+}\sigma_{+}$ and $\sigma_{+}\sigma_{-}$ configurations, which can be used to test some possible two-phonon assignments.
	
	\begin{figure}[htbp]
		\includegraphics[width=1.0\columnwidth]{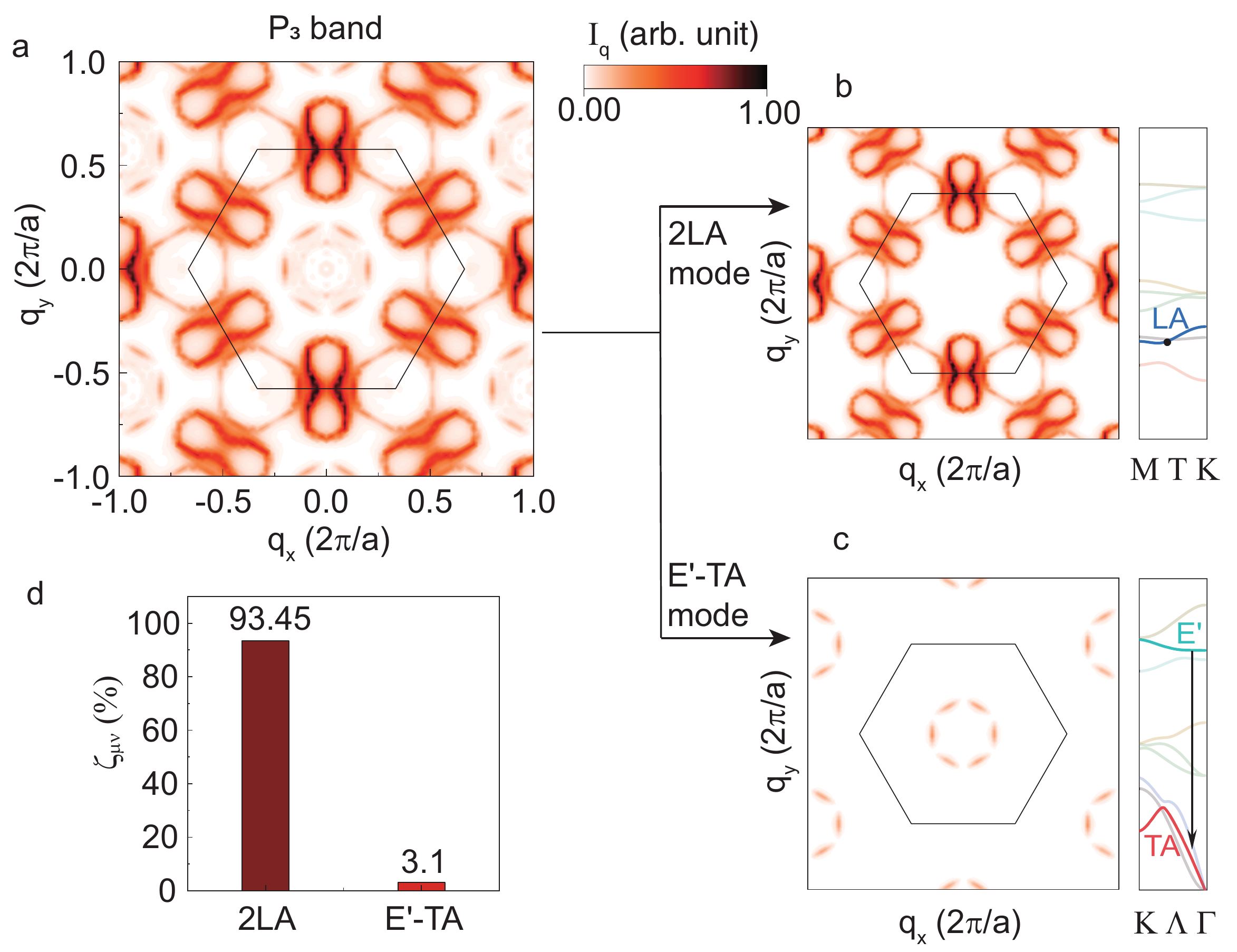}
		\caption{(Color online) The assignment of the P$_3$ DRR band. (a) Raman intensity as a function of $q$ in the whole BZ. The main two-phonon combinations are 2LA (overtone) and E$^{\prime}$ - TA (subtraction) whose Raman intensities as a function of $q$ are given in (b) and (c), respectively. (d) The percentage of the 2LA and E$^{\prime}$ - TA modes.}
		\label{Fig3}
	\end{figure}
	Next, let us assign each DRR mode. In Fig.~\ref{Fig3}, we take the P$_3$ mode at laser excitation energy 1.96 eV as an example, which was previously assigned to be 2LA(M) overtone mode~\cite{Guo15}. Basically one specific Raman shift (205 cm$^{-1}$ here) with peak intensity is chosen, then we calculate the scattering cross section as a function of $q$, I$_{q}$ (see Eq.~\ref{eq3}) from all possible two-phonon modes including overtone, combination, subtraction ones. The Raman intensity I$_{q}$ is illustrated in Fig.~\ref{Fig3}(a), and the two dominating DRR modes are derived to be the 2LA (overtone) and E$^{\prime}$ - TA (subtraction) as indicated on the right panel of Fig.~\ref{Fig3}(a). The percentage $\zeta_{\mu\nu}$ (See Eq.~\ref{eq4}) in Fig.~\ref{Fig3}(d) shows that the 2LA dominates over the E$^{\prime}$ - TA. Such an assignment is consistent with the previous group theory analysis made in the Appendix Table~\ref{Table2}.
	\begin{table}[htbp]
		\centering
		\caption{A comparison of assignment for second-order Raman modes P$_i$ (i = 1, 2, ..., 6) between the
			previous work~\cite{Guo15} and the current work for E$_L$ = 1.96 eV. The label (new) denotes new assignments
			by the present work. The percentage of Raman intensity for each DRR mode is given.}
		\scalebox{1.0}
		{
			\begin{tabular}{p{0.9cm}<{\centering}|p{2.90cm}|p{2.90cm}|p{1.20cm}}
				\hline
				\hline
				{mode }& {Ref.[\onlinecite{Guo15,Caramazza2018}]} & {this work}  &{$\zeta_{\mu\nu}$}\\
				\hline
				{P$_1$}& {2TA(M)}                   & {2TA(M)}             &{9.2\%} \\
				{     }& {{E$^{\prime}$(M) - LA(M)}}     &{{E$^{\prime}$ - LA}}      &{84.9\%} \\
				\hline
				{P$_2$}& {{E$^{\prime}$(M) - TA(M)}}     & {{E$^{\prime}$ - TA}}     &{18.4\%} \\
				{     }& {   }                      & {LA + TA (new)}      &{55.2\%} \\
				{     }& {   }                      & {2LA (new)}          &{13.9\%} \\
				{     }& {   }                      & {E$^{\prime}$ - LA (new)} &{7.6\%} \\
				\hline
				{P$_3$}& {E$^{\prime\prime}$(M) + TA(M)}     &{ NA in $\bar{\textrm{Z}}$(XX)Z} &{ }\\
				{     }& {{2LA(M)}}                &{2LA(M)}               &{93.5\%}\\
				{     }& { }                       &{E$^{\prime}$ - TA  (new)}  &{3.1\%}\\
				\hline
				{P$_4$}& {{A$^{\prime}_{1}$(M) + LA(M)}} & {A$^{\prime}_{1}$ + LA}    & {80.0\% }\\
				{ }    & { }                        & {E$^{\prime}$  + LA (new)} & {7.7\%}\\
				\hline
				{P$_5$}& {E$^{\prime}$(M) + TA(M)}       & {E$^{\prime}$  + TA}       & {27.4\% }\\
				{  }   & {    }                     & {E$^{\prime}$  + LA (new)} & {58.2\% }\\
				\hline
				{P$_6$}& {E$^{\prime}$(M) + LA(M)}       & {E$^{\prime}$  + LA}       & {98.7\% }\\
				\hline
				\hline
			\end{tabular}
			\label{Table1}
		}
	\end{table}
	
	Further, not only the mode assignment (2LA), but also the most probable phonon vectors $q$ can be obtained from the calculated Raman spectra as shown in Fig.~\ref{Fig3}(a), which is distributed around the $M$ points with a "8" shape, and agrees reasonably with the results derived from equi-energy contour lines of band structure by Guo \textit{et al.} (see Fig. 2c in Ref.~\cite{Guo15}. However, Guo \textit{et al.}~\cite{Guo15} in theory and Caramazza \textit{et al.}~\cite{Caramazza2018} in experiment also assigned the P$_3$ band partly to E$^{\prime\prime}$(M) + TA(M), which is absent at M from our current analysis, merely because we have used the backscattering configuration ($\bar{\textrm{Z}}$(XX)Z) in our calculation, in which E$^{\prime\prime}$ is not Raman-active.
	
	The data of the remaining DRR modes P$_i$ (i = 1, 2, 4-6) is given in Table~\ref{Table1} and also in the Appendix. The mode assignments are compared directly with the previous work by Guo \textit{et al.}~\cite{Guo15} and Caramazza \textit{et al.}~\cite{Caramazza2018}, as listed in Table~\ref{Table1}. From Table~\ref{Table1}, we confirm that most assignments in this work for all the six DRR modes are consistent with that made by Guo \textit{et al.}~\cite{Guo15} and Caramazza \textit{et al.}~\cite{Caramazza2018}. Nevertheless, we also have non-zero $\zeta$ for other combination modes which appear in the other regions of the $q$ space. One of the advantages of the present calculation is to be able to supply the fraction of each assignment when multiple scattering channels are possible, for example, E$^{\prime}$ - LA ($\zeta_{\mu\nu} \approx$85\%) is much more important than 2TA ($\zeta_{\mu\nu} \approx$9\%) for P$_1$, both P$_3$ and P$_6$ modes have one dominating assignment (2LA and E$^{\prime}$ + LA, respectively), all of which can not be extracted merely from group theoretical analysis. Moreover, some additional assignments can be unveiled from this work, for example, for P$_2$, besides E$^{\prime}$ - TA, it has LA + TA, 2LA, and E$^{\prime}$ - LA, in which LA + TA ($\zeta_{\mu\nu} \approx$55\%) is even more important than E$^{\prime}$ - TA ($\zeta_{\mu\nu} \approx$18\%).
	
	\begin{figure*}[htbp]
		\includegraphics[width=1.8\columnwidth]{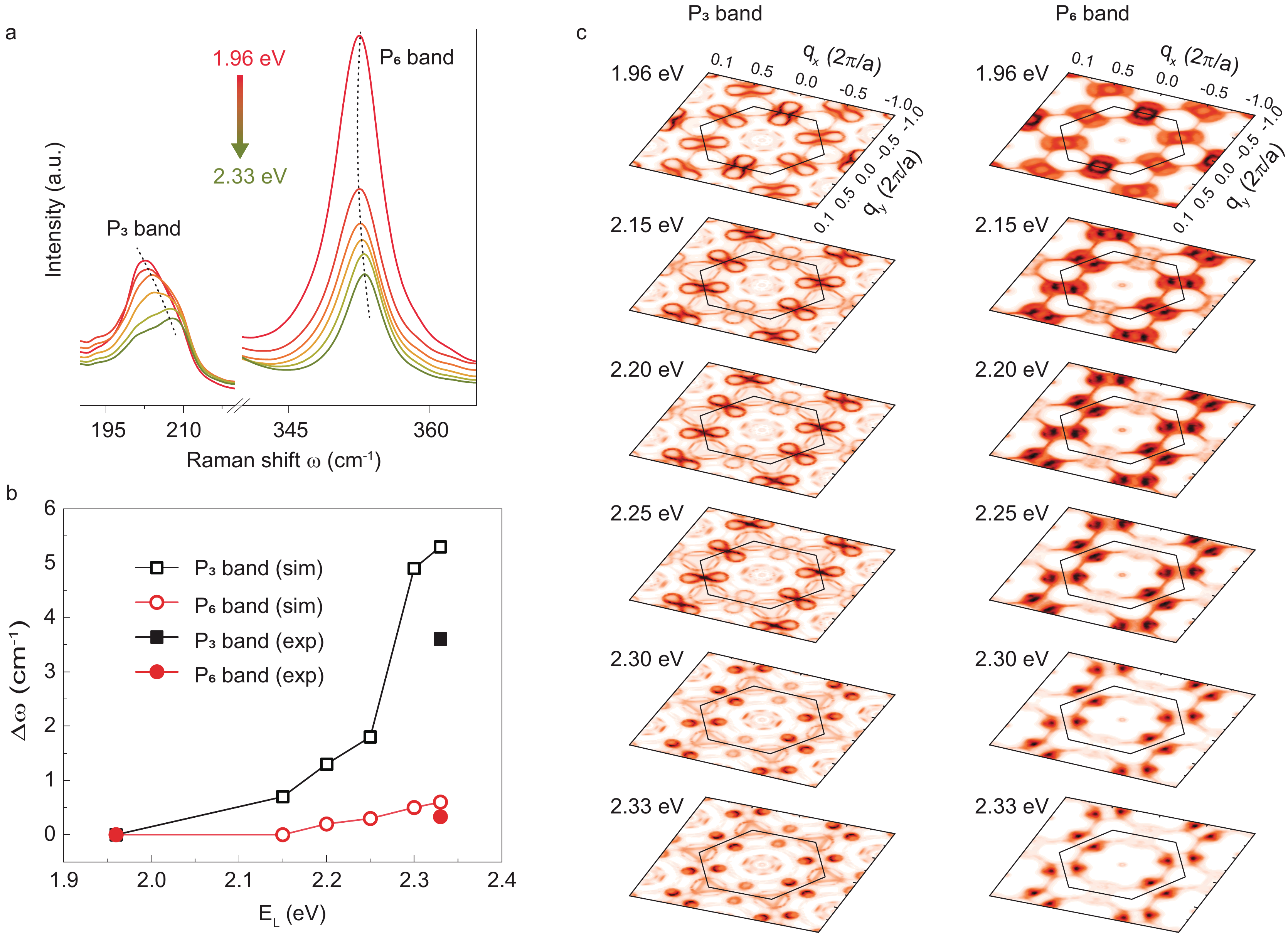}
		\caption{(Color online) The dispersion of DRR modes on laser energy E$_L$. The E$_L$ dependence of (a) Raman profile, (b) Raman shift, and (c) $q$-resolved Raman intensity of the P$_3$ and P$_6$ bands.}
		\label{Fig4}
	\end{figure*}
	What is more, as seen from the Raman intensity as a function of $q$ and laser excitation energy E$_L$ in Fig.~\ref{Fig4}(c) and also in the Appendix Fig.~\ref{SI-Fig1}-\ref{SI-Fig5}, Raman intensity can arise from general $q$ points with low symmetries, the $q$ points which contribute most to total Raman intensity appear either at or between the high-symmetry points in BZ, which justifies the previous theoretical treatment only on the high-symmetry points~\cite{Sekine84,Stacy85,Chen74,Sourisseau89,Sourisseau1991,Feldman96,Frey99,Windom11,Li12,Chakraborty13,Terrones14,Carvalho2017,Livneh2015,Guo15,Caramazza2018}. However, there is one DRR peak at around 300 cm$^{-1}$ which has not been explored before due to its overlap with the Raman peak of Silicon. If looking more carefully into the experimental Raman spectra in Fig.~\ref{Fig1}(a), one can see a satellite peak on the right of the Silicon peak. This peak, which we designate as P$_7$ band, is more visible from our simulated Raman spectra, and can be assigned to be an overtone mode, namely 2E$^{\prime\prime}$, at $q$ close to the $K$ point along $\Gamma$ and $K$ points, as analyzed in Appendix Fig.~\ref{SI-Fig6}.
	
	Since the second-order Raman spectra are dispersive as a function of excitation energy E$_L$, we further substantiate our mode assignment by showing the dispersion of two typical DRR bands P$_3$ and P$_6$ with E$_L$ in Fig.~\ref{Fig4}. From the E$_L$ dependent band profile (Fig.~\ref{Fig4}(a)) and especially peak shift (with respect to the frequency at 1.96 eV) (Fig.~\ref{Fig4}(b)), we can see the dispersive features in the P$_3$ and P$_6$ DRR bands, both of which are blue-shifted with increasing E$_L$, with $\Delta\omega$/$\Delta$E$_L$ $\sim$ 25.6 and 3.33 cm$^{-1}$/eV for P$_3$ and P$_6$, respectively. The simulation results (in open squares and circles) agree reasonably well with the experimental data (in solid squares and circles). The dispersion difference between the P$_3$ and P$_6$ bands arises mainly from E$_L$ dependent phonon wave vector $q$. As shown in Fig.~\ref{Fig4}(c), with increasing E$_L$ from 1.96 eV to 2.33 eV, the $q$ changes from between $M$ and $K$, via $M$ point, to between $M$ and $\Gamma$ points for the P$_3$ band. In contrast, the $q$ for the P$_6$ band keeps almost unchanged with increasing E$_L$. The P$_3$ band dispersion can also be seen from the phonon dispersion relation in Fig.~\ref{Fig2}(b), the band segment of the LA band around the $M$ point that is assigned to P$_3$ (Table~\ref{Table1}) is rather dispersive.
	
	\begin{figure}[htbp]
		\includegraphics[width=1.0\columnwidth]{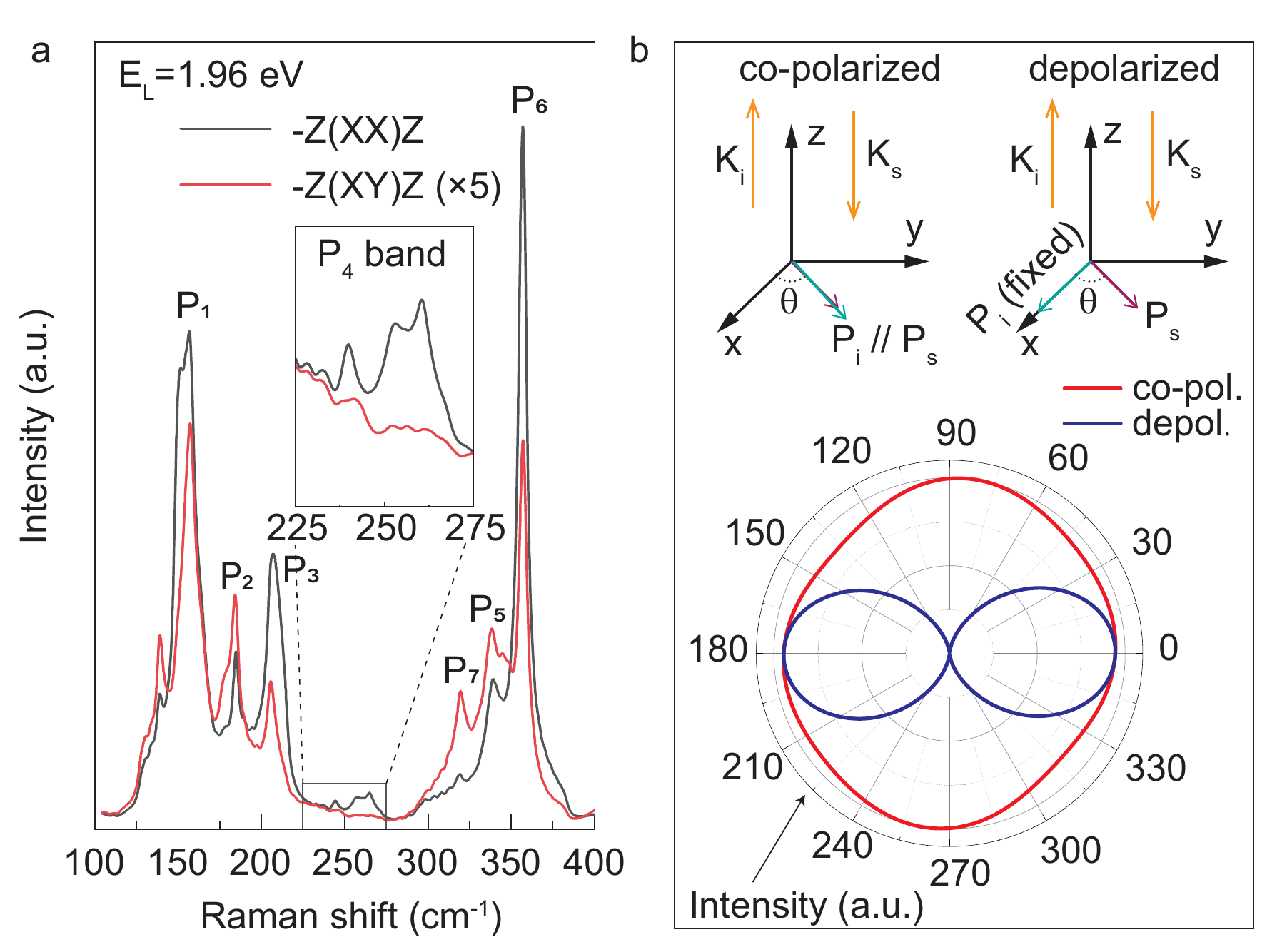}
		\caption{(Color online) The polarization selectivity of the DRR modes at E$_L$ = 1.96 eV. (a) The linearly polarized Raman spectra in the parallel ($\bar{\textrm{Z}}$(XX)Z) and perpendicular ($\bar{\textrm{Z}}$(XX)Z) geometries in black and red, respectively. The intensity of the ($\bar{\textrm{Z}}$(XX)Z) configuration is multiplied by 5. The P$_4$ band is enlarged in inset. (b) The polar figure of Raman intensity for both co-polarized and depolarized set-ups.}
		\label{Fig5}
	\end{figure}
	Finally, let us show the laser polarization dependence of the DRR modes. From Table~\ref{Table1}, we can see that most of the DRR modes have E$^{\prime}$ involved, except for the P$_4$ band. The P$_4$ band has two most probable assignments A$^{\prime}_{1}$ + LA and E$^{\prime}$ + LA, but the former of which has a much bigger fraction ($\zeta_{\mu\nu} \approx$80\%) than the latter one ($\zeta_{\mu\nu} \approx$7.7\%), suggesting a potentially strong linear polarization dependence. This is indeed the case, as indicated in Fig.~\ref{Fig5}(a). Figure~\ref{Fig5}(a) gives the calculated Raman spectra in both the parallel ($\bar{\textrm{Z}}$(XX)Z) and perpendicular ($\bar{\textrm{Z}}$(XX)Z) geometries. The intensities in the two geometries are nonzero and almost equal for all the DRR modes discussed here, except for the P$_4$ band, which has zero intensity in the perpendicular geometries. We also explore the angle dependence of Raman intensity of the P$_4$ band. The co-polarized and depolarized geometries are both used and set-ups are schematized in Fig.~\ref{Fig5}(b). The polar plot of both geometries is consistent with the above analysis. The isotropic polar data are seen for the co-polarization geometry and comes from the in-plane isotropy of monolayer MoTe$_2$, while the depolarized geometry that the incident and scattered laser polarizations have a relative angle $\theta$ gives a strong anisotropy ($\approx$ cos$^2\theta$) arising from A$^{\prime}_1$, consistent with the group theoretical analysis in the Appendix Table~\ref{Table2}.
	
	\section{conclusion}
	In summary, we have calculated the second-order Raman spectra of MoTe$_2$ monolayer, based on first-principles density functional calculation and time-dependent perturbation theory. The non-empirical treatment of electron-phonon interaction is performed, which allows us to quantify the contribution of all possible two-phonon combinations to the double resonance Raman modes in every single phonon wave vector. The P$_i$ (i = 1, 2, ..., 7) band assignments, which are consistent with the DRR selective rule constructed based on group theory, show some additional origin of two phonon modes which was not found in the previous studies. The polarization dependence of the DRR modes are also investigated, which should be observed experimentally. This study facilitates a deeper understanding of the electron-phonon interaction and the second-order Raman process in TMDC systems.
	
	\begin{acknowledgments}
		We sincerely acknowledge Prof. Vitto Zheng Han for helping prepare the high-quality figures. This project is supported by the National Natural Science Foundation of China (52031014, 51702146) and the National Key R\&D Program of China (2017YFA0206301). H.G. acknowledges Department of Education of Liaoning Province Grants No. LJKZ0391. L.Z. acknowledges the National Key Basic Research Program of China (2021YFA1401400) and the National Natural Science Foundation of China (52103344). L.T. acknowledges the National Natural Science Foundation of China (21974004). R.S. acknowledges MEXT Grants No. JP18H01810. The simulation work was carried out at National Supercomputer Center in Tianjin, China, and the calculations were performed on TianHe-1(A).
	\end{acknowledgments}

\clearpage
	
\appendix

\onecolumngrid

\section{Raman selection rule analysis of two phonons}
\vspace{-7mm}
\begin{table*}[htbp]
	\centering
	\caption{Raman selection rule analysis of two phonons of MoTe$_2$ with different symmetries based on group theory. The light blue and brown color mark the Raman active and inactive phonon combinations, respectively.}
	\label{Table2}
	\scalebox{0.8}
	{
		\begin{tabular}{|c|c||c|c|c||c|c|c|}
			\hhline{--||---||---}
			\rowcolor[HTML]{BDEEBC}
			Product & Reduction & Product & \multicolumn{2}{c||}{\cellcolor[HTML]{BDEEBC}Reduction} & Product & \multicolumn{2}{c|}{\cellcolor[HTML]{BDEEBC}Reduction} \\ \hhline{--||---||---}
			\rowcolor[HTML]{BDEEBC}
			$\textrm{D}_\textrm{3h}$ & $\textrm{D}_\textrm{3h}$ & $\textrm{C}_\textrm{3h}$ & $\textrm{D}_\textrm{3h}$ & $\textrm{C}_\textrm{3h}$ & $\textrm{C}_\textrm{2v}$ & $\textrm{D}_\textrm{3h}$ & $\textrm{C}_\textrm{2v}$  \\ \hhline{--||---||---}
			\rowcolor[HTML]{AFB1D3}
			$\textrm{A}^{\prime}_1\times \textrm{A}^{\prime}_1$ & $\textrm{A}^{\prime}_1$ & $\textrm{A}^{\prime}\times \textrm{A}^{\prime}$ & $\textrm{A}^{\prime}_1+\textrm{A}^{\prime}_2$ & $\textrm{A}^{\prime}$ & $\textrm{A}_1\times \textrm{A}_1$ & $\textrm{A}^{\prime}_1+\textrm{E}^{\prime}$ & $\textrm{A}_1+\textrm{B}_1$ \\ \hhline{--||---||---}
			\rowcolor[HTML]{FFCE93}
			$\textrm{A}^{\prime}_1\times \textrm{A}^{\prime}_2$ & $\textrm{A}^{\prime}_2$ & $\textrm{A}^{\prime}\times \textrm{A}^{\prime\prime}$ &  $\textrm{A}^{\prime\prime}_1+\textrm{A}^{\prime\prime}_2$ & $\textrm{A}^{\prime\prime}$ &  $\textrm{A}_1\times \textrm{A}_2$ & $\textrm{A}^{\prime\prime}_1+\textrm{E}^{\prime\prime}$ & $\textrm{A}_2+\textrm{B}_2$ \\ \hhline{--||---||---}
			\rowcolor[HTML]{AFB1D3}
			\cellcolor[HTML]{FFCE93}$\textrm{A}^{\prime}_1\times \textrm{A}^{\prime\prime}_1$ & \cellcolor[HTML]{FFCE93}$\textrm{A}^{\prime\prime}_1$ & $\textrm{A}^{\prime}\times {}^{1}\textrm{E}^{\prime}$ & $\textrm{E}^{\prime}$ & ${}^{1}\textrm{E}^{\prime}$ & $\textrm{A}_1\times \textrm{B}_1$ & $\textrm{A}^{\prime}_2+\textrm{E}^{\prime}$ & $\textrm{A}_1+\textrm{B}_1$ \\ \hhline{--||---||---}
			\rowcolor[HTML]{FFCE93}
			$\textrm{A}^{\prime}_1\times \textrm{A}^{\prime\prime}_2$ & $\textrm{A}^{\prime\prime}_2$ & \cellcolor[HTML]{AFB1D3}$\textrm{A}^{\prime}\times {}^{2}\textrm{E}^{\prime}$ & \cellcolor[HTML]{AFB1D3}$\textrm{E}^{\prime}$ & \cellcolor[HTML]{AFB1D3}${}^{2}\textrm{E}^{\prime}$ & $\textrm{A}_1\times \textrm{B}_2$ & $\textrm{A}^{\prime\prime}_2+\textrm{E}^{\prime\prime}$ & $\textrm{A}_2+\textrm{B}_2$ \\ \hhline{--||---||---}
			\rowcolor[HTML]{AFB1D3}
			$\textrm{A}^{\prime}_1\times \textrm{E}^{\prime}$ & $\textrm{E}^{\prime}$ & \cellcolor[HTML]{FFCE93}$\textrm{A}^{\prime}\times {}^{1}\textrm{E}^{\prime\prime}$ & \cellcolor[HTML]{FFCE93}$\textrm{E}^{\prime\prime}$ & \cellcolor[HTML]{FFCE93}${}^{1}\textrm{E}^{\prime\prime}$ & $\textrm{A}_2\times \textrm{A}_2$ & $\textrm{A}^{\prime}_1+\textrm{E}^{\prime}$ & $\textrm{A}_1+\textrm{B}_1$ \\ \hhline{--||---||---}
			\rowcolor[HTML]{FFCE93}
			$\textrm{A}^{\prime}_1\times \textrm{E}^{\prime\prime}$ & $\textrm{E}^{\prime\prime}$ & $\textrm{A}^{\prime}\times {}^{2}\textrm{E}^{\prime\prime}$ & $\textrm{E}^{\prime\prime}$ & ${}^{2}\textrm{E}^{\prime\prime}$ & $\textrm{A}_2\times \textrm{B}_1$ & $\textrm{A}^{\prime\prime}_2+\textrm{E}^{\prime\prime}$ & $\textrm{A}_2+\textrm{B}_2$ \\ \hhline{--||---||---}
			\rowcolor[HTML]{AFB1D3}
			$\textrm{A}^{\prime}_2\times \textrm{A}^{\prime}_2$ & $\textrm{A}^{\prime}_1$ & $\textrm{A}^{\prime\prime}\times\textrm{A}^{\prime\prime}$ & $\textrm{A}^{\prime}_1+\textrm{A}^{\prime}_2$ & $\textrm{A}^{\prime}$ &  $\textrm{A}_2\times \textrm{B}_2$ & $\textrm{A}^{\prime}_2+\textrm{E}^{\prime}$ & $\textrm{A}_1+\textrm{B}_1$ \\ \hhline{--||---||---}
			\rowcolor[HTML]{FFCE93}
			$\textrm{A}^{\prime}_2\times \textrm{A}^{\prime\prime}_1$ & $\textrm{A}^{\prime\prime}_2$ & $\textrm{A}^{\prime\prime}\times {}^{1}\textrm{E}^{\prime}$ & $\textrm{E}^{\prime\prime}$ & ${}^{1}\textrm{E}^{\prime\prime}$ & \cellcolor[HTML]{AFB1D3}$\textrm{B}_1\times \textrm{B}_1$ & \cellcolor[HTML]{AFB1D3}$\textrm{A}^{\prime}_1+\textrm{E}^{\prime}$ & \cellcolor[HTML]{AFB1D3}$\textrm{A}_1+\textrm{B}_1$ \\ \hhline{--||---||---}
			\rowcolor[HTML]{FFCE93}
			$\textrm{A}^{\prime}_2\times \textrm{A}^{\prime\prime}_2$ & $\textrm{A}^{\prime\prime}_1$ & $\textrm{A}^{\prime\prime}\times {}^{2}\textrm{E}^{\prime}$ & $\textrm{E}^{\prime\prime}$ & ${}^{2}\textrm{E}^{\prime\prime}$ & $\textrm{B}_1\times \textrm{B}_2$ & $\textrm{A}^{\prime\prime}_1+\textrm{E}^{\prime\prime}$ & $\textrm{A}_2+\textrm{B}_2$ \\ \hhline{--||---||---}
			\rowcolor[HTML]{AFB1D3}
			$\textrm{A}^{\prime}_2\times \textrm{E}^{\prime}$ & $\textrm{E}^{\prime}$ & $\textrm{A}^{\prime\prime}\times {}^{1}\textrm{E}^{\prime\prime}$ & $\textrm{E}^{\prime}$ & ${}^{1}\textrm{E}^{\prime}$ & $\textrm{B}_2\times \textrm{B}_2$ & $\textrm{A}^{\prime}_1+\textrm{E}^{\prime}$ & $\textrm{A}_1+\textrm{B}_1$ \\ \hhline{--||---||---}
			\cellcolor[HTML]{FFCE93}$\textrm{A}^{\prime}_2\times \textrm{E}^{\prime\prime}$ & \cellcolor[HTML]{FFCE93}$\textrm{E}^{\prime\prime}$ & \cellcolor[HTML]{AFB1D3}$\textrm{A}^{\prime\prime}\times {}^{2}\textrm{E}^{\prime\prime}$ & \cellcolor[HTML]{AFB1D3}$\textrm{E}^{\prime}$ & \cellcolor[HTML]{AFB1D3}${}^{2}\textrm{E}^{\prime}$ & \cellcolor[HTML]{BDEEBC}Product & \multicolumn{2}{c|}{\cellcolor[HTML]{BDEEBC}Reduction} \\ \hhline{--||---||---}
			\rowcolor[HTML]{AFB1D3}
			$\textrm{A}^{\prime\prime}_1\times \textrm{A}^{\prime\prime}_1$ & $\textrm{A}^{\prime}_1$ & ${}^{1}\textrm{E}^{\prime}\times {}^{1}\textrm{E}^{\prime}$ & $\textrm{A}^{\prime}_1+\textrm{A}^{\prime}_2$ & ${}^{2}\textrm{E}^{\prime}$ & \cellcolor[HTML]{BDEEBC} & {\cellcolor[HTML]{BDEEBC}}$\textrm{D}_\textrm{3h}$ & {\cellcolor[HTML]{BDEEBC}}$\textrm{C}_\textrm{3h}$ \\ \hhline{--||---||~--}
			\cellcolor[HTML]{FFCE93}$\textrm{A}^{\prime\prime}_1\times \textrm{A}^{\prime\prime}_2$ & \cellcolor[HTML]{FFCE93}$\textrm{A}^{\prime}_2$ & \cellcolor[HTML]{AFB1D3}${}^{1}\textrm{E}^{\prime}\times {}^{2}\textrm{E}^{\prime}$ & \cellcolor[HTML]{AFB1D3}$\textrm{E}^{\prime}$ & \cellcolor[HTML]{AFB1D3}$\textrm{A}^{\prime}$ & \multirow{-2}{*}{\cellcolor[HTML]{BDEEBC}$\textrm{C}_\textrm{s}$} & {\cellcolor[HTML]{BDEEBC}}$\textrm{C}_\textrm{2v}$ & {\cellcolor[HTML]{BDEEBC}}$\textrm{C}_\textrm{s}$ \\ \hhline{--||---||---}
			\rowcolor[HTML]{FFCE93}
			$\textrm{A}^{\prime\prime}_1\times \textrm{E}^{\prime}$ & $\textrm{E}^{\prime\prime}$ & ${}^{1}\textrm{E}^{\prime}\times {}^{1}\textrm{E}^{\prime\prime}$ & $\textrm{A}^{\prime\prime}_1+\textrm{A}^{\prime\prime}_2$ & ${}^{2}\textrm{E}^{\prime\prime}$ & \cellcolor[HTML]{AFB1D3} & \cellcolor[HTML]{AFB1D3}$\textrm{A}^{\prime}_1+\textrm{A}^{\prime}_2+2\textrm{E}^{\prime}$ & \cellcolor[HTML]{AFB1D3}$\textrm{A}^{\prime}+{}^{1}\textrm{E}^{\prime}+{}^{2}\textrm{E}^{\prime}$ \\ \hhline{--||---||~--}
			\rowcolor[HTML]{AFB1D3}
			$\textrm{A}^{\prime\prime}_1\times \textrm{E}^{\prime\prime}$ & $\textrm{E}^{\prime}$ & \cellcolor[HTML]{FFCE93}${}^{1}\textrm{E}^{\prime}\times {}^{2}\textrm{E}^{\prime\prime}$ & \cellcolor[HTML]{FFCE93}$\textrm{E}^{\prime\prime}$ & \cellcolor[HTML]{FFCE93}$\textrm{A}^{\prime\prime}$ & \multirow{-2}{*}{\cellcolor[HTML]{AFB1D3}$\textrm{A}^{\prime}\times \textrm{A}^{\prime}$} & $2\textrm{A}_1+2\textrm{B}_1$ & $2\textrm{A}^{\prime}$ \\ \hhline{--||---||---}
			\rowcolor[HTML]{AFB1D3}
			$\textrm{A}^{\prime\prime}_2\times \textrm{A}^{\prime\prime}_2$ & $\textrm{A}^{\prime}_1$ & ${}^{2}\textrm{E}^{\prime}\times {}^{2}\textrm{E}^{\prime}$ & $\textrm{A}^{\prime}_1+\textrm{A}^{\prime}_2$ & ${}^{1}\textrm{E}^{\prime}$ & \cellcolor[HTML]{FFCE93} & \cellcolor[HTML]{FFCE93}$\textrm{A}^{\prime\prime}_1+\textrm{A}^{\prime\prime}_2+2\textrm{E}^{\prime\prime}$ & \cellcolor[HTML]{FFCE93}$\textrm{A}^{\prime\prime}+{}^{1}\textrm{E}^{\prime\prime}+{}^{2}\textrm{E}^{\prime\prime}$ \\ \hhline{--||---||~--}
			\rowcolor[HTML]{FFCE93}
			$\textrm{A}^{\prime\prime}_2\times \textrm{E}^{\prime}$ & $\textrm{E}^{\prime\prime}$ & ${}^{2}\textrm{E}^{\prime}\times {}^{1}\textrm{E}^{\prime\prime}$ & $\textrm{E}^{\prime\prime}$ & $\textrm{A}^{\prime\prime}$ & \multirow{-2}{*}{\cellcolor[HTML]{FFCE93}$\textrm{A}^{\prime}\times \textrm{A}^{\prime\prime}$} & $2\textrm{A}_2+2\textrm{B}_2$ & $2\textrm{A}^{\prime\prime}$ \\ \hhline{--||---||---}
			\rowcolor[HTML]{AFB1D3}
			$\textrm{A}^{\prime\prime}_2\times \textrm{E}^{\prime\prime}$ & $\textrm{E}^{\prime}$ & \cellcolor[HTML]{FFCE93}${}^{2}\textrm{E}^{\prime}\times {}^{2}\textrm{E}^{\prime\prime}$ & \cellcolor[HTML]{FFCE93}$\textrm{A}^{\prime\prime}_1+\textrm{A}^{\prime}_2$ & \cellcolor[HTML]{FFCE93}${}^{1}\textrm{E}^{\prime\prime}$ & \cellcolor[HTML]{AFB1D3} & $\textrm{A}^{\prime}_1+\textrm{A}^{\prime}_2+2\textrm{E}^{\prime}$ & $\textrm{A}^{\prime}+{}^{1}\textrm{E}^{\prime}+{}^{2}\textrm{E}^{\prime}$ \\ \hhline{--||---||~--}
			\rowcolor[HTML]{AFB1D3}
			$\textrm{E}^{\prime}\times \textrm{E}^{\prime}$ &  $\textrm{A}^{\prime}_1+\textrm{A}^{\prime}_2+\textrm{E}^{\prime}$ & ${}^{1}\textrm{E}^{\prime\prime}\times {}^{1}\textrm{E}^{\prime\prime}$ & $\textrm{A}^{\prime}_1+\textrm{A}^{\prime}_2$ & ${}^{2}\textrm{E}^{\prime}$ & \multirow{-2}{*}{\cellcolor[HTML]{AFB1D3}$\textrm{A}^{\prime\prime}\times \textrm{A}^{\prime\prime}$} & $2\textrm{A}_1+2\textrm{B}_1$ & $2\textrm{A}^{\prime}$ \\ \hhline{--||---||---}
			\rowcolor[HTML]{AFB1D3}
			\cellcolor[HTML]{FFCE93}$\textrm{E}^{\prime}\times \textrm{E}^{\prime\prime}$ & \cellcolor[HTML]{FFCE93}$\textrm{A}^{\prime\prime}_1+\textrm{A}^{\prime\prime}_2+\textrm{E}^{\prime\prime}$ & ${}^{1}\textrm{E}^{\prime\prime}\times {}^{2}\textrm{E}^{\prime\prime}$ & $\textrm{E}^{\prime}$ & $\textrm{A}^{\prime}$ & \multicolumn{3}{c|}{\cellcolor[HTML]{AFB1D3}Raman active} \\ \hhline{--||---||---}
			\rowcolor[HTML]{AFB1D3}
			$\textrm{E}^{\prime\prime}\times \textrm{E}^{\prime\prime}$ & $\textrm{A}^{\prime}_1+\textrm{A}^{\prime}_2+\textrm{E}^{\prime}$ & ${}^{2}\textrm{E}^{\prime\prime}\times {}^{2}\textrm{E}^{\prime\prime}$ & $\textrm{A}^{\prime}_1+\textrm{A}^{\prime}_2$ & ${}^{1}\textrm{E}^{\prime}$ & \multicolumn{3}{c|}{\cellcolor[HTML]{FFCE93}Raman inactive} \\ \hhline{--||---||---}
		\end{tabular}
	}
\end{table*}
\vspace{-3mm}
\section{The double resonance Raman modes P$_1$-P$_2$, P$_4$-P$_7$}
\twocolumngrid

\begin{figure}[htbp]
	\includegraphics[width=0.85\columnwidth]{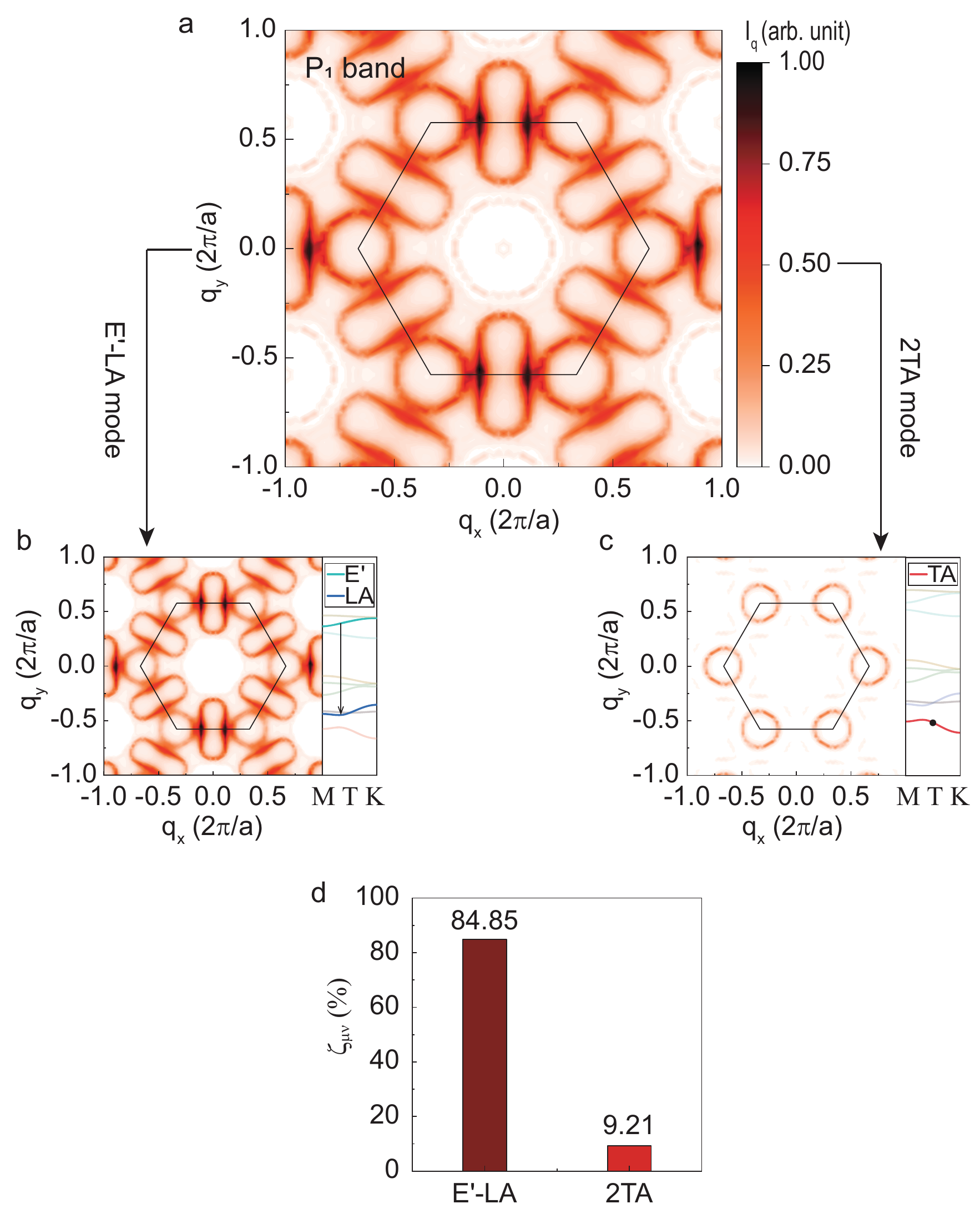}
	\caption{(Color online) The P$_1$ double resonance Raman mode. (a) $q$-resolved Raman intensity in the whole BZ. The main two-phonon combinations are E$^{\prime}$ - LA (subtraction) and 2TA (overtone) whose $q$-dependent Raman intensities are given in (b) and (c), respectively. (d) The percentage of the E$^{\prime}$ - LA and 2TA modes.}
	\label{SI-Fig1}
\end{figure}

\begin{figure}[htbp]
	\includegraphics[width=0.85\columnwidth]{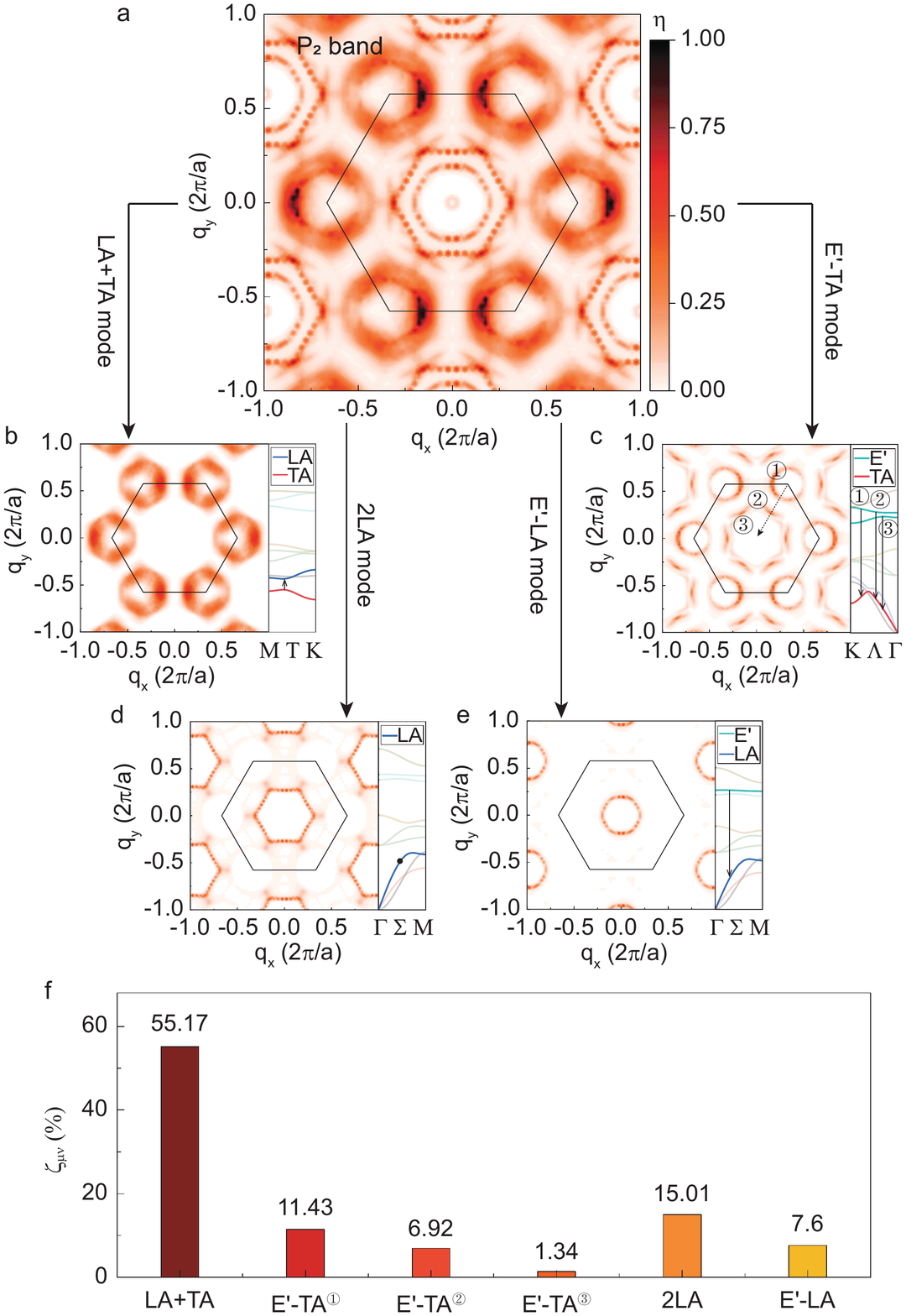}
	\caption{(Color online) The P$_2$ double resonance Raman mode. (a) $q$-resolved Raman intensity in the whole BZ. The main two-phonon combinations are LA + TA,  E$^{\prime}$ - TA, 2LA, E$^{\prime}$ - LA whose $q$-dependent Raman intensities are given in (b-e), respectively. (f) The percentage of all the assigned modes.}
	\label{SI-Fig2}
\end{figure}

\begin{figure}[htbp]
	\includegraphics[width=0.85\columnwidth]{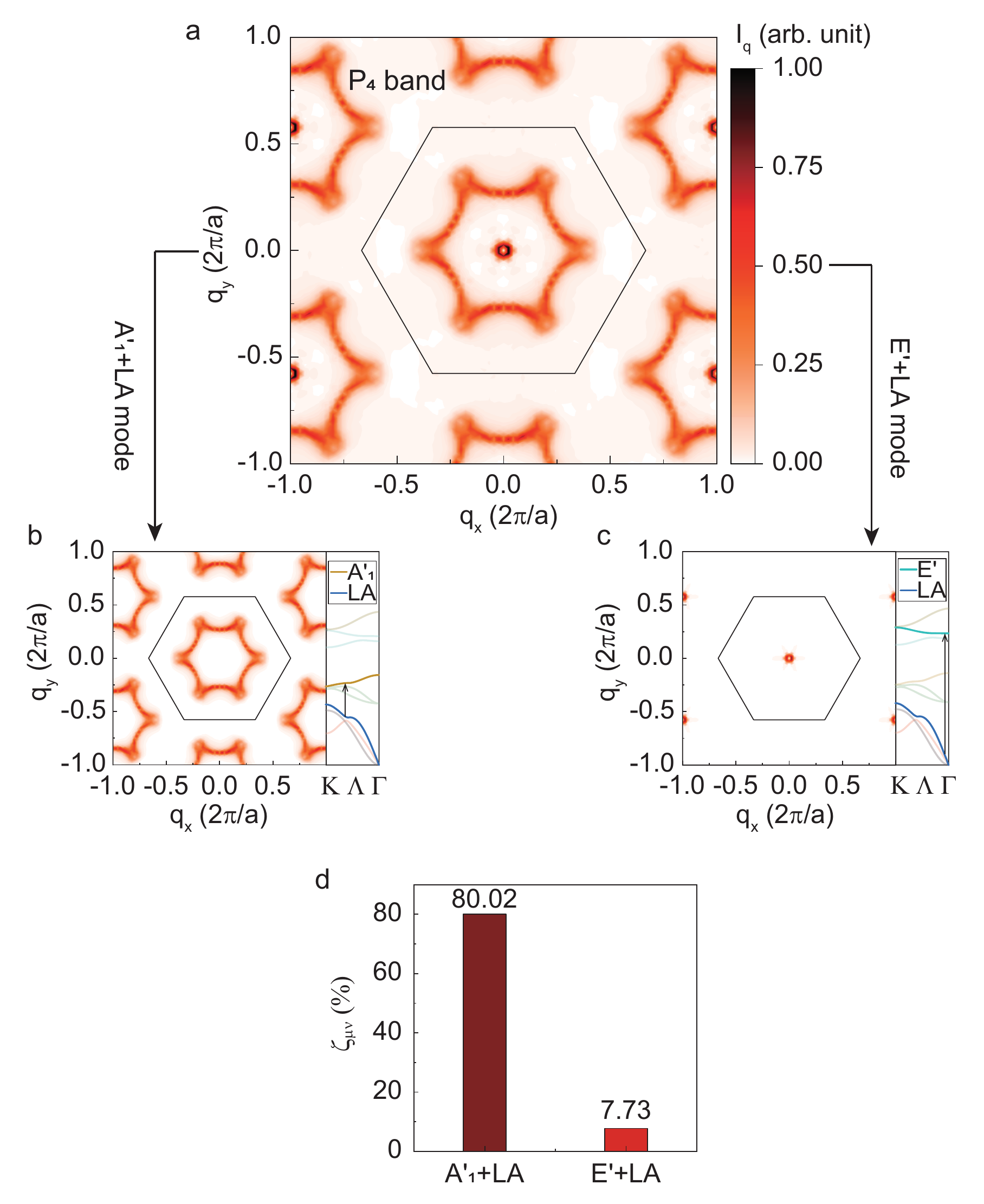}
	\caption{(Color online) The P$_4$ double resonance Raman mode. (a) $q$-resolved Raman intensity in the whole BZ. The main two-phonon combinations are A$^{\prime}_1$ + LA and E$^{\prime}$ + LA whose
		$q$-dependent Raman intensities are given in (b) and (c), respectively. (d) The percentage of the A$^{\prime}_1$ + LA and E$^{\prime}$ + LA modes.}
	\label{SI-Fig3}
\end{figure}

\begin{figure}[htbp]
	\includegraphics[width=0.85\columnwidth]{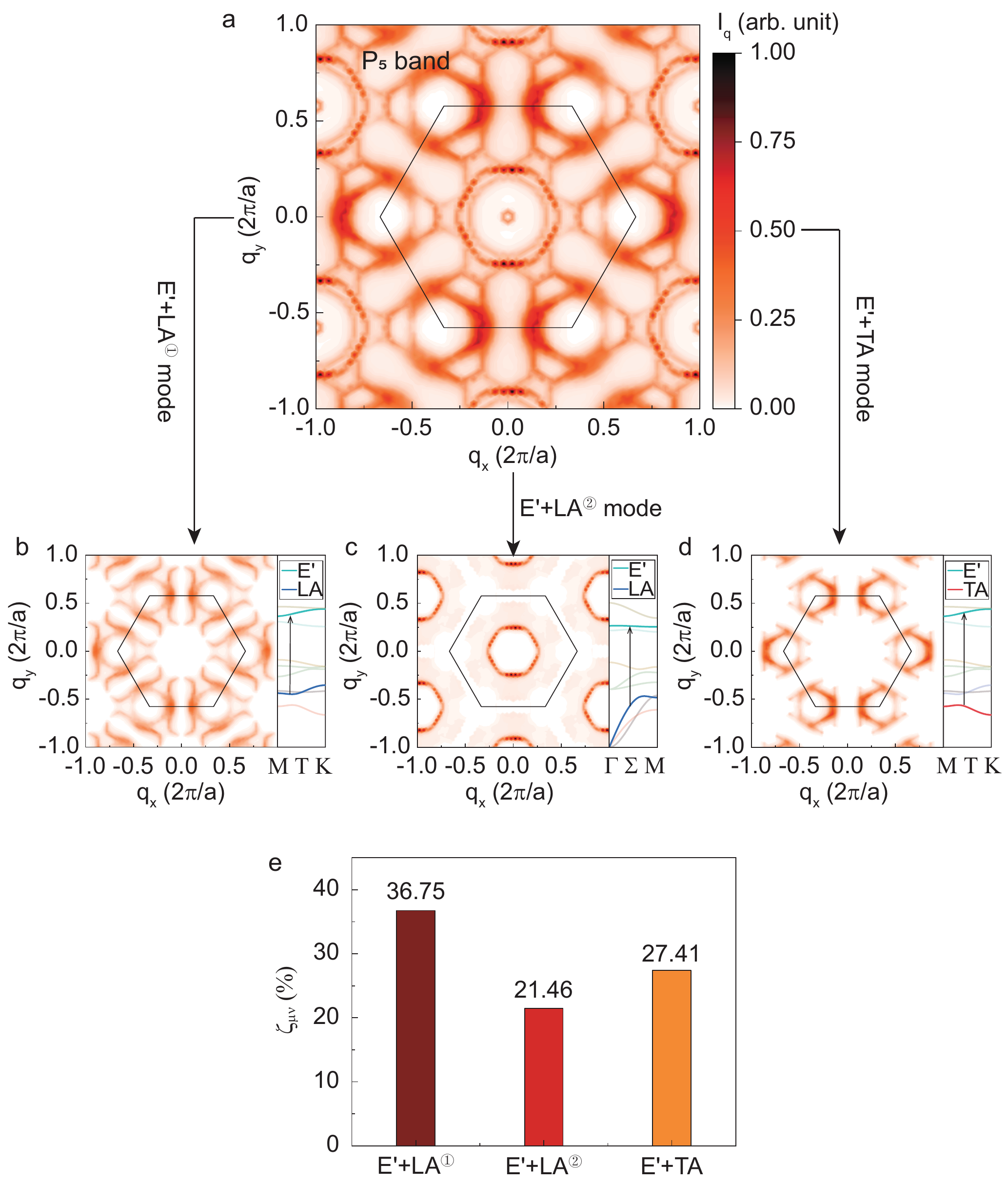}
	\caption{(Color online) The P$_5$ double resonance Raman mode. (a) $q$-resolved Raman intensity in the whole BZ. The main two-phonon combinations are E$^{\prime}$ + LA and E$^{\prime}$ + TA
		whose $q$-dependent Raman intensities are given in (b-d), respectively. (e) The percentage of the E$^{\prime}$ + LA and E$^{\prime}$ + TA modes.}
	\label{SI-Fig4}
\end{figure}

\begin{figure}[htbp]
	\includegraphics[width=0.85\columnwidth]{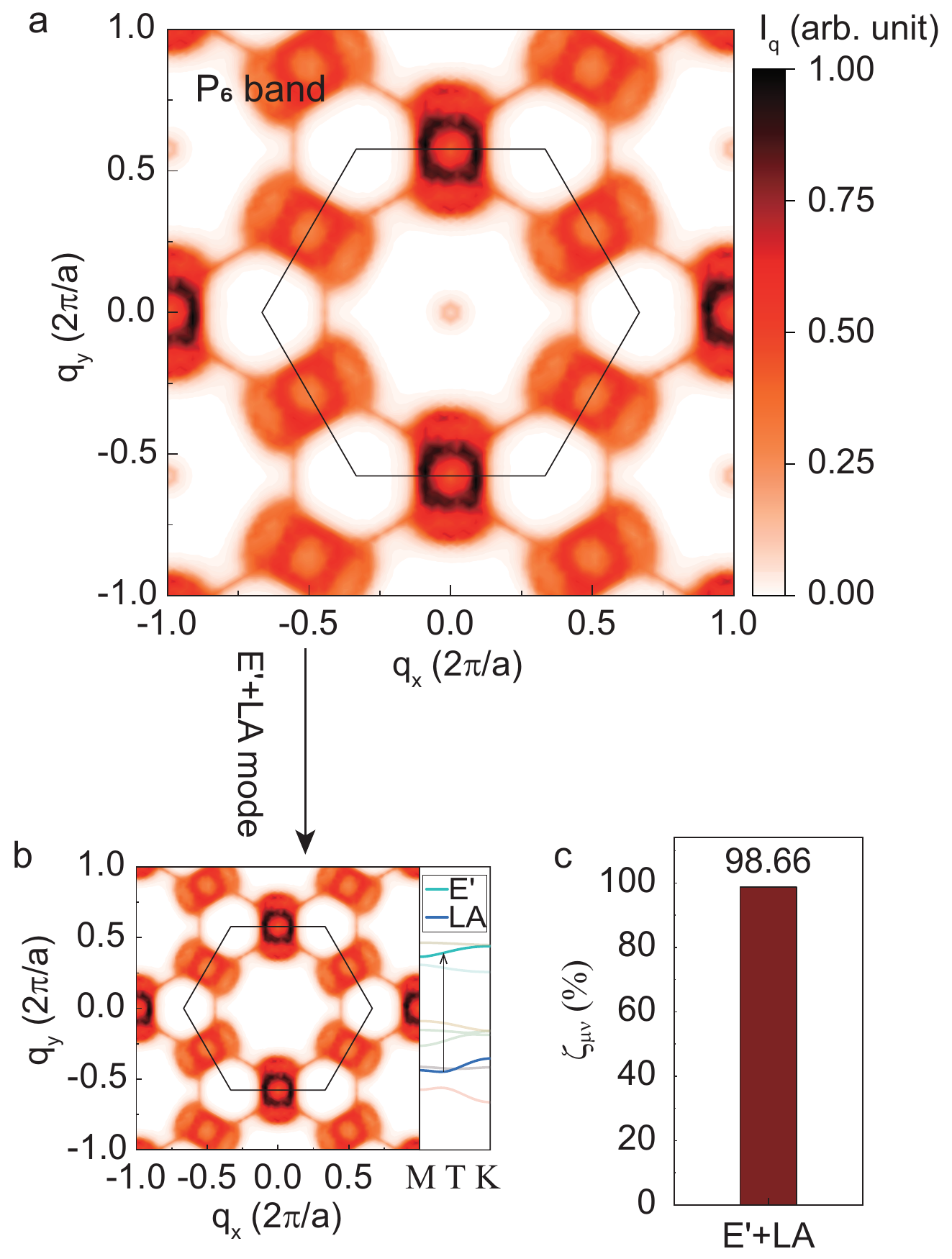}
	\caption{(Color online) The P$_6$ double resonance Raman mode. (a) $q$-resolved Raman intensity in the whole BZ. The dominating two-phonon combination is from E$^{\prime}$ + LA whose $q$-dependent
		Raman intensities is given in (b). (c) The percentage of the E$^{\prime}$ + LA.}
	\label{SI-Fig5}
\end{figure}

\begin{figure}[htbp]
	\includegraphics[width=0.85\columnwidth]{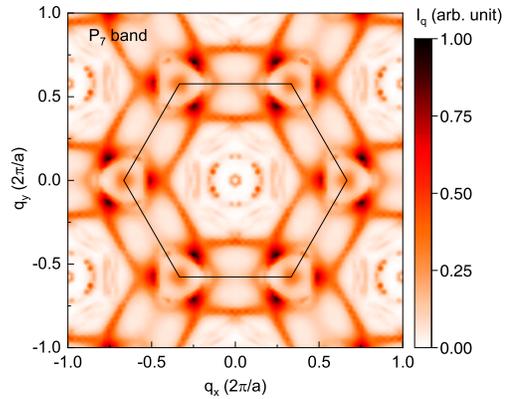}
	\caption{(Color online) The P$_7$ double resonance Raman mode. The dominating combination is from 2E$^{\prime\prime}$ whose $q$ is distributed close the $K$ point between $\Gamma$ and $K$ points.}
	\label{SI-Fig6}
\end{figure}

\clearpage


\begin{thebibliography}{47}%
\makeatletter
\providecommand \@ifxundefined [1]{%
 \@ifx{#1\undefined}
}%
\providecommand \@ifnum [1]{%
 \ifnum #1\expandafter \@firstoftwo
 \else \expandafter \@secondoftwo
 \fi
}%
\providecommand \@ifx [1]{%
 \ifx #1\expandafter \@firstoftwo
 \else \expandafter \@secondoftwo
 \fi
}%
\providecommand \natexlab [1]{#1}%
\providecommand \enquote  [1]{``#1''}%
\providecommand \bibnamefont  [1]{#1}%
\providecommand \bibfnamefont [1]{#1}%
\providecommand \citenamefont [1]{#1}%
\providecommand \href@noop [0]{\@secondoftwo}%
\providecommand \href [0]{\begingroup \@sanitize@url \@href}%
\providecommand \@href[1]{\@@startlink{#1}\@@href}%
\providecommand \@@href[1]{\endgroup#1\@@endlink}%
\providecommand \@sanitize@url [0]{\catcode `\\12\catcode `\$12\catcode
  `\&12\catcode `\#12\catcode `\^12\catcode `\_12\catcode `\%12\relax}%
\providecommand \@@startlink[1]{}%
\providecommand \@@endlink[0]{}%
\providecommand \url  [0]{\begingroup\@sanitize@url \@url }%
\providecommand \@url [1]{\endgroup\@href {#1}{\urlprefix }}%
\providecommand \urlprefix  [0]{URL }%
\providecommand \Eprint [0]{\href }%
\providecommand \doibase [0]{https://doi.org/}%
\providecommand \selectlanguage [0]{\@gobble}%
\providecommand \bibinfo  [0]{\@secondoftwo}%
\providecommand \bibfield  [0]{\@secondoftwo}%
\providecommand \translation [1]{[#1]}%
\providecommand \BibitemOpen [0]{}%
\providecommand \bibitemStop [0]{}%
\providecommand \bibitemNoStop [0]{.\EOS\space}%
\providecommand \EOS [0]{\spacefactor3000\relax}%
\providecommand \BibitemShut  [1]{\csname bibitem#1\endcsname}%
\let\auto@bib@innerbib\@empty
\bibitem [{\citenamefont {Measson}\ \emph {et~al.}(2014)\citenamefont
  {Measson}, \citenamefont {Gallais}, \citenamefont {Cazayous}, \citenamefont
  {Clair}, \citenamefont {Rodiere}, \citenamefont {Cario},\ and\ \citenamefont
  {Sacuto}}]{Measson2014}%
  \BibitemOpen
  \bibfield  {author} {\bibinfo {author} {\bibfnamefont {M.~A.}\ \bibnamefont
  {Measson}}, \bibinfo {author} {\bibfnamefont {Y.}~\bibnamefont {Gallais}},
  \bibinfo {author} {\bibfnamefont {M.}~\bibnamefont {Cazayous}}, \bibinfo
  {author} {\bibfnamefont {B.}~\bibnamefont {Clair}}, \bibinfo {author}
  {\bibfnamefont {P.}~\bibnamefont {Rodiere}}, \bibinfo {author} {\bibfnamefont
  {L.}~\bibnamefont {Cario}},\ and\ \bibinfo {author} {\bibfnamefont
  {A.}~\bibnamefont {Sacuto}},\ }\href@noop {} {\bibfield  {journal} {\bibinfo
  {journal} {Phys. Rev. B}\ }\textbf {\bibinfo {volume} {89}},\ \bibinfo
  {pages} {060503(R)} (\bibinfo {year} {2014})}\BibitemShut {NoStop}%
\bibitem [{\citenamefont {Kung}\ \emph {et~al.}(2017)\citenamefont {Kung},
  \citenamefont {Maiti}, \citenamefont {Wang}, \citenamefont {Cheong},
  \citenamefont {Maslov},\ and\ \citenamefont {Blumberg}}]{Kung2017}%
  \BibitemOpen
  \bibfield  {author} {\bibinfo {author} {\bibfnamefont {H.-H.}\ \bibnamefont
  {Kung}}, \bibinfo {author} {\bibfnamefont {S.}~\bibnamefont {Maiti}},
  \bibinfo {author} {\bibfnamefont {X.}~\bibnamefont {Wang}}, \bibinfo {author}
  {\bibfnamefont {S.-W.}\ \bibnamefont {Cheong}}, \bibinfo {author}
  {\bibfnamefont {D.~L.}\ \bibnamefont {Maslov}},\ and\ \bibinfo {author}
  {\bibfnamefont {G.}~\bibnamefont {Blumberg}},\ }\href@noop {} {\bibfield
  {journal} {\bibinfo  {journal} {Phys. Rev. Lett.}\ }\textbf {\bibinfo
  {volume} {119}},\ \bibinfo {pages} {136802} (\bibinfo {year}
  {2017})}\BibitemShut {NoStop}%
\bibitem [{\citenamefont {Tenne}\ \emph {et~al.}(2006)\citenamefont {Tenne},
  \citenamefont {Bruchhausen}, \citenamefont {Lanzillotti-Kimura},
  \citenamefont {Fainstein}, \citenamefont {Katiyar}, \citenamefont
  {Cantarero}, \citenamefont {Soukiassian}, \citenamefont {Vaithyanathan},
  \citenamefont {Haeni}, \citenamefont {Tian}, \citenamefont {Schlom},
  \citenamefont {Choi}, \citenamefont {Kim}, \citenamefont {Eom}, \citenamefont
  {Sun}, \citenamefont {Pan}, \citenamefont {Li}, \citenamefont {Chen},
  \citenamefont {Jia}, \citenamefont {Nakhmanson}, \citenamefont {Rabe},\ and\
  \citenamefont {Xi}}]{Tenne2006}%
  \BibitemOpen
  \bibfield  {author} {\bibinfo {author} {\bibfnamefont {D.~A.}\ \bibnamefont
  {Tenne}}, \bibinfo {author} {\bibfnamefont {A.}~\bibnamefont {Bruchhausen}},
  \bibinfo {author} {\bibfnamefont {N.~D.}\ \bibnamefont {Lanzillotti-Kimura}},
  \bibinfo {author} {\bibfnamefont {A.}~\bibnamefont {Fainstein}}, \bibinfo
  {author} {\bibfnamefont {R.~S.}\ \bibnamefont {Katiyar}}, \bibinfo {author}
  {\bibfnamefont {A.}~\bibnamefont {Cantarero}}, \bibinfo {author}
  {\bibfnamefont {A.}~\bibnamefont {Soukiassian}}, \bibinfo {author}
  {\bibfnamefont {V.}~\bibnamefont {Vaithyanathan}}, \bibinfo {author}
  {\bibfnamefont {J.~H.}\ \bibnamefont {Haeni}}, \bibinfo {author}
  {\bibfnamefont {W.}~\bibnamefont {Tian}}, \bibinfo {author} {\bibfnamefont
  {D.~G.}\ \bibnamefont {Schlom}}, \bibinfo {author} {\bibfnamefont {K.~J.}\
  \bibnamefont {Choi}}, \bibinfo {author} {\bibfnamefont {D.~M.}\ \bibnamefont
  {Kim}}, \bibinfo {author} {\bibfnamefont {C.~B.}\ \bibnamefont {Eom}},
  \bibinfo {author} {\bibfnamefont {H.~P.}\ \bibnamefont {Sun}}, \bibinfo
  {author} {\bibfnamefont {X.~Q.}\ \bibnamefont {Pan}}, \bibinfo {author}
  {\bibfnamefont {Y.~L.}\ \bibnamefont {Li}}, \bibinfo {author} {\bibfnamefont
  {L.~Q.}\ \bibnamefont {Chen}}, \bibinfo {author} {\bibfnamefont {Q.~X.}\
  \bibnamefont {Jia}}, \bibinfo {author} {\bibfnamefont {S.~M.}\ \bibnamefont
  {Nakhmanson}}, \bibinfo {author} {\bibfnamefont {K.~M.}\ \bibnamefont
  {Rabe}},\ and\ \bibinfo {author} {\bibfnamefont {X.~X.}\ \bibnamefont {Xi}},\
  }\href@noop {} {\bibfield  {journal} {\bibinfo  {journal} {Science}\ }\textbf
  {\bibinfo {volume} {313}},\ \bibinfo {pages} {1614} (\bibinfo {year}
  {2006})}\BibitemShut {NoStop}%
\bibitem [{\citenamefont {Huang}\ \emph {et~al.}(2020)\citenamefont {Huang},
  \citenamefont {Cenker}, \citenamefont {Zhang}, \citenamefont {Ray},
  \citenamefont {Song}, \citenamefont {Taniguchi}, \citenamefont {Watanabe},
  \citenamefont {McGuire}, \citenamefont {Xiao},\ and\ \citenamefont
  {Xu}}]{Huang2020}%
  \BibitemOpen
  \bibfield  {author} {\bibinfo {author} {\bibfnamefont {B.}~\bibnamefont
  {Huang}}, \bibinfo {author} {\bibfnamefont {J.}~\bibnamefont {Cenker}},
  \bibinfo {author} {\bibfnamefont {X.}~\bibnamefont {Zhang}}, \bibinfo
  {author} {\bibfnamefont {E.~L.}\ \bibnamefont {Ray}}, \bibinfo {author}
  {\bibfnamefont {T.}~\bibnamefont {Song}}, \bibinfo {author} {\bibfnamefont
  {T.}~\bibnamefont {Taniguchi}}, \bibinfo {author} {\bibfnamefont
  {K.}~\bibnamefont {Watanabe}}, \bibinfo {author} {\bibfnamefont {M.~A.}\
  \bibnamefont {McGuire}}, \bibinfo {author} {\bibfnamefont {D.}~\bibnamefont
  {Xiao}},\ and\ \bibinfo {author} {\bibfnamefont {X.}~\bibnamefont {Xu}},\
  }\href@noop {} {\bibfield  {journal} {\bibinfo  {journal} {Nat.
  Nanotechnol.}\ }\textbf {\bibinfo {volume} {15}},\ \bibinfo {pages} {212}
  (\bibinfo {year} {2020})}\BibitemShut {NoStop}%
\bibitem [{\citenamefont {Kim}\ \emph {et~al.}(2019)\citenamefont {Kim},
  \citenamefont {Lim}, \citenamefont {Lee}, \citenamefont {Lee}, \citenamefont
  {Kim}, \citenamefont {Park}, \citenamefont {Jeon}, \citenamefont {Park},
  \citenamefont {Park},\ and\ \citenamefont {Cheong}}]{Kim2019}%
  \BibitemOpen
  \bibfield  {author} {\bibinfo {author} {\bibfnamefont {K.}~\bibnamefont
  {Kim}}, \bibinfo {author} {\bibfnamefont {S.~Y.}\ \bibnamefont {Lim}},
  \bibinfo {author} {\bibfnamefont {J.-U.}\ \bibnamefont {Lee}}, \bibinfo
  {author} {\bibfnamefont {S.}~\bibnamefont {Lee}}, \bibinfo {author}
  {\bibfnamefont {T.~Y.}\ \bibnamefont {Kim}}, \bibinfo {author} {\bibfnamefont
  {K.}~\bibnamefont {Park}}, \bibinfo {author} {\bibfnamefont {G.~S.}\
  \bibnamefont {Jeon}}, \bibinfo {author} {\bibfnamefont {C.-H.}\ \bibnamefont
  {Park}}, \bibinfo {author} {\bibfnamefont {J.-G.}\ \bibnamefont {Park}},\
  and\ \bibinfo {author} {\bibfnamefont {H.}~\bibnamefont {Cheong}},\
  }\href@noop {} {\bibfield  {journal} {\bibinfo  {journal} {Nat. Commun.}\
  }\textbf {\bibinfo {volume} {10}},\ \bibinfo {pages} {345} (\bibinfo {year}
  {2019})}\BibitemShut {NoStop}%
\bibitem [{\citenamefont {Miranda}\ \emph {et~al.}(2017)\citenamefont
  {Miranda}, \citenamefont {Reichardt}, \citenamefont {Froehlicher},
  \citenamefont {Molina-Sanchez}, \citenamefont {Berciaud},\ and\ \citenamefont
  {Wirtz}}]{Miranda2017}%
  \BibitemOpen
  \bibfield  {author} {\bibinfo {author} {\bibfnamefont {H.~P.~C.}\
  \bibnamefont {Miranda}}, \bibinfo {author} {\bibfnamefont {S.}~\bibnamefont
  {Reichardt}}, \bibinfo {author} {\bibfnamefont {G.}~\bibnamefont
  {Froehlicher}}, \bibinfo {author} {\bibfnamefont {A.}~\bibnamefont
  {Molina-Sanchez}}, \bibinfo {author} {\bibfnamefont {S.}~\bibnamefont
  {Berciaud}},\ and\ \bibinfo {author} {\bibfnamefont {L.}~\bibnamefont
  {Wirtz}},\ }\href@noop {} {\bibfield  {journal} {\bibinfo  {journal} {Nano
  Lett.}\ }\textbf {\bibinfo {volume} {17}},\ \bibinfo {pages} {2381} (\bibinfo
  {year} {2017})}\BibitemShut {NoStop}%
\bibitem [{\citenamefont {Zhang}\ \emph {et~al.}(2022)\citenamefont {Zhang},
  \citenamefont {Huang}, \citenamefont {Yu}, \citenamefont {Wang},
  \citenamefont {Yang}, \citenamefont {Zhang}, \citenamefont {Tong},\ and\
  \citenamefont {Zhang}}]{Zhang2022}%
  \BibitemOpen
  \bibfield  {author} {\bibinfo {author} {\bibfnamefont {S.}~\bibnamefont
  {Zhang}}, \bibinfo {author} {\bibfnamefont {J.}~\bibnamefont {Huang}},
  \bibinfo {author} {\bibfnamefont {Y.}~\bibnamefont {Yu}}, \bibinfo {author}
  {\bibfnamefont {S.}~\bibnamefont {Wang}}, \bibinfo {author} {\bibfnamefont
  {T.}~\bibnamefont {Yang}}, \bibinfo {author} {\bibfnamefont {Z.}~\bibnamefont
  {Zhang}}, \bibinfo {author} {\bibfnamefont {L.}~\bibnamefont {Tong}},\ and\
  \bibinfo {author} {\bibfnamefont {J.}~\bibnamefont {Zhang}},\ }\href
  {https://doi.org/10.1038/s41467-022-28877-6} {\bibfield  {journal} {\bibinfo
  {journal} {Nature Communications}\ }\textbf {\bibinfo {volume} {13}},\
  \bibinfo {pages} {1254} (\bibinfo {year} {2022})}\BibitemShut {NoStop}%
\bibitem [{\citenamefont {Tatsumi}\ and\ \citenamefont
  {Saito}(2018)}]{Tatsumi2017}%
  \BibitemOpen
  \bibfield  {author} {\bibinfo {author} {\bibfnamefont {Y.}~\bibnamefont
  {Tatsumi}}\ and\ \bibinfo {author} {\bibfnamefont {R.}~\bibnamefont
  {Saito}},\ }\href@noop {} {\bibfield  {journal} {\bibinfo  {journal} {Phys.
  Rev. B}\ }\textbf {\bibinfo {volume} {97}},\ \bibinfo {pages} {115407}
  (\bibinfo {year} {2018})}\BibitemShut {NoStop}%
\bibitem [{\citenamefont {Chen}\ \emph {et~al.}(2016)\citenamefont {Chen},
  \citenamefont {Goldstein}, \citenamefont {Venkataraman}, \citenamefont
  {Ramasubramaniam},\ and\ \citenamefont {Yan}}]{Chen2016}%
  \BibitemOpen
  \bibfield  {author} {\bibinfo {author} {\bibfnamefont {S.-Y.}\ \bibnamefont
  {Chen}}, \bibinfo {author} {\bibfnamefont {T.}~\bibnamefont {Goldstein}},
  \bibinfo {author} {\bibfnamefont {D.}~\bibnamefont {Venkataraman}}, \bibinfo
  {author} {\bibfnamefont {A.}~\bibnamefont {Ramasubramaniam}},\ and\ \bibinfo
  {author} {\bibfnamefont {J.}~\bibnamefont {Yan}},\ }\href@noop {} {\bibfield
  {journal} {\bibinfo  {journal} {Nano Letters}\ }\textbf {\bibinfo {volume}
  {16}},\ \bibinfo {pages} {5852} (\bibinfo {year} {2016})}\BibitemShut
  {NoStop}%
\bibitem [{\citenamefont {Zhang}\ \emph {et~al.}(2016)\citenamefont {Zhang},
  \citenamefont {Bao}, \citenamefont {Gu}, \citenamefont {Ren}, \citenamefont
  {Zhang}, \citenamefont {Deng}, \citenamefont {Wu}, \citenamefont {Li},
  \citenamefont {Feng},\ and\ \citenamefont {Zhou}}]{Zhang2016}%
  \BibitemOpen
  \bibfield  {author} {\bibinfo {author} {\bibfnamefont {K.}~\bibnamefont
  {Zhang}}, \bibinfo {author} {\bibfnamefont {C.}~\bibnamefont {Bao}}, \bibinfo
  {author} {\bibfnamefont {Q.}~\bibnamefont {Gu}}, \bibinfo {author}
  {\bibfnamefont {X.}~\bibnamefont {Ren}}, \bibinfo {author} {\bibfnamefont
  {H.}~\bibnamefont {Zhang}}, \bibinfo {author} {\bibfnamefont
  {K.}~\bibnamefont {Deng}}, \bibinfo {author} {\bibfnamefont {Y.}~\bibnamefont
  {Wu}}, \bibinfo {author} {\bibfnamefont {Y.}~\bibnamefont {Li}}, \bibinfo
  {author} {\bibfnamefont {J.}~\bibnamefont {Feng}},\ and\ \bibinfo {author}
  {\bibfnamefont {S.}~\bibnamefont {Zhou}},\ }\href
  {https://doi.org/10.1038/ncomms13552} {\bibfield  {journal} {\bibinfo
  {journal} {Nat. Commun.}\ }\textbf {\bibinfo {volume} {7}},\ \bibinfo {pages}
  {13552} (\bibinfo {year} {2016})}\BibitemShut {NoStop}%
\bibitem [{\citenamefont {Liu}\ \emph {et~al.}(2018)\citenamefont {Liu},
  \citenamefont {Gu}, \citenamefont {Peng}, \citenamefont {Qi}, \citenamefont
  {Zhang}, \citenamefont {Zhang}, \citenamefont {Ma}, \citenamefont {Zhu},
  \citenamefont {Tong}, \citenamefont {Feng}, \citenamefont {Liu},\ and\
  \citenamefont {Chen}}]{Liu2018}%
  \BibitemOpen
  \bibfield  {author} {\bibinfo {author} {\bibfnamefont {Y.}~\bibnamefont
  {Liu}}, \bibinfo {author} {\bibfnamefont {Q.}~\bibnamefont {Gu}}, \bibinfo
  {author} {\bibfnamefont {Y.}~\bibnamefont {Peng}}, \bibinfo {author}
  {\bibfnamefont {S.}~\bibnamefont {Qi}}, \bibinfo {author} {\bibfnamefont
  {N.}~\bibnamefont {Zhang}}, \bibinfo {author} {\bibfnamefont
  {Y.}~\bibnamefont {Zhang}}, \bibinfo {author} {\bibfnamefont
  {X.}~\bibnamefont {Ma}}, \bibinfo {author} {\bibfnamefont {R.}~\bibnamefont
  {Zhu}}, \bibinfo {author} {\bibfnamefont {L.}~\bibnamefont {Tong}}, \bibinfo
  {author} {\bibfnamefont {J.}~\bibnamefont {Feng}}, \bibinfo {author}
  {\bibfnamefont {Z.}~\bibnamefont {Liu}},\ and\ \bibinfo {author}
  {\bibfnamefont {J.-H.}\ \bibnamefont {Chen}},\ }\href@noop {} {\bibfield
  {journal} {\bibinfo  {journal} {Adv. Mater.}\ }\textbf {\bibinfo {volume}
  {30}},\ \bibinfo {pages} {1706402} (\bibinfo {year} {2018})}\BibitemShut
  {NoStop}%
\bibitem [{\citenamefont {Zhang}\ \emph {et~al.}(2020)\citenamefont {Zhang},
  \citenamefont {Pang}, \citenamefont {Wang}, \citenamefont {Han},
  \citenamefont {Shang}, \citenamefont {Hung}, \citenamefont {Nugraha},
  \citenamefont {Liu}, \citenamefont {Li}, \citenamefont {Saito},\ and\
  \citenamefont {Huang}}]{Zhang2020}%
  \BibitemOpen
  \bibfield  {author} {\bibinfo {author} {\bibfnamefont {K.}~\bibnamefont
  {Zhang}}, \bibinfo {author} {\bibfnamefont {X.}~\bibnamefont {Pang}},
  \bibinfo {author} {\bibfnamefont {T.}~\bibnamefont {Wang}}, \bibinfo {author}
  {\bibfnamefont {F.}~\bibnamefont {Han}}, \bibinfo {author} {\bibfnamefont
  {S.-L.}\ \bibnamefont {Shang}}, \bibinfo {author} {\bibfnamefont {N.~T.}\
  \bibnamefont {Hung}}, \bibinfo {author} {\bibfnamefont {A.~R.~T.}\
  \bibnamefont {Nugraha}}, \bibinfo {author} {\bibfnamefont {Z.-K.}\
  \bibnamefont {Liu}}, \bibinfo {author} {\bibfnamefont {M.}~\bibnamefont
  {Li}}, \bibinfo {author} {\bibfnamefont {R.}~\bibnamefont {Saito}},\ and\
  \bibinfo {author} {\bibfnamefont {S.}~\bibnamefont {Huang}},\ }\href@noop {}
  {\bibfield  {journal} {\bibinfo  {journal} {Phys. Rev. B}\ }\textbf {\bibinfo
  {volume} {101}},\ \bibinfo {pages} {014308} (\bibinfo {year}
  {2020})}\BibitemShut {NoStop}%
\bibitem [{\citenamefont {Wakabayashi}\ \emph {et~al.}(1975)\citenamefont
  {Wakabayashi}, \citenamefont {Smith},\ and\ \citenamefont
  {Nicklow}}]{Wakabayashi75}%
  \BibitemOpen
  \bibfield  {author} {\bibinfo {author} {\bibfnamefont {N.}~\bibnamefont
  {Wakabayashi}}, \bibinfo {author} {\bibfnamefont {H.~G.}\ \bibnamefont
  {Smith}},\ and\ \bibinfo {author} {\bibfnamefont {R.~M.}\ \bibnamefont
  {Nicklow}},\ }\href@noop {} {\bibfield  {journal} {\bibinfo  {journal} {Phys.
  Rev. B}\ }\textbf {\bibinfo {volume} {12}},\ \bibinfo {pages} {659} (\bibinfo
  {year} {1975})}\BibitemShut {NoStop}%
\bibitem [{\citenamefont {Sourisseau}\ \emph {et~al.}(1991)\citenamefont
  {Sourisseau}, \citenamefont {Cruege}, \citenamefont {Fouassier},\ and\
  \citenamefont {Alba}}]{Sourisseau1991}%
  \BibitemOpen
  \bibfield  {author} {\bibinfo {author} {\bibfnamefont {C.}~\bibnamefont
  {Sourisseau}}, \bibinfo {author} {\bibfnamefont {F.}~\bibnamefont {Cruege}},
  \bibinfo {author} {\bibfnamefont {M.}~\bibnamefont {Fouassier}},\ and\
  \bibinfo {author} {\bibfnamefont {M.}~\bibnamefont {Alba}},\ }\href@noop {}
  {\bibfield  {journal} {\bibinfo  {journal} {Chem. Phys.}\ }\textbf {\bibinfo
  {volume} {150}},\ \bibinfo {pages} {281} (\bibinfo {year}
  {1991})}\BibitemShut {NoStop}%
\bibitem [{\citenamefont {Sekine}\ \emph {et~al.}(1984)\citenamefont {Sekine},
  \citenamefont {Uchinokura}, \citenamefont {Nakashizu}, \citenamefont
  {Matsuura},\ and\ \citenamefont {Yoshizaki}}]{Sekine84}%
  \BibitemOpen
  \bibfield  {author} {\bibinfo {author} {\bibfnamefont {T.}~\bibnamefont
  {Sekine}}, \bibinfo {author} {\bibfnamefont {K.}~\bibnamefont {Uchinokura}},
  \bibinfo {author} {\bibfnamefont {T.}~\bibnamefont {Nakashizu}}, \bibinfo
  {author} {\bibfnamefont {E.}~\bibnamefont {Matsuura}},\ and\ \bibinfo
  {author} {\bibfnamefont {R.}~\bibnamefont {Yoshizaki}},\ }\href@noop {}
  {\bibfield  {journal} {\bibinfo  {journal} {J. Phys. Soc. Jpn.}\ }\textbf
  {\bibinfo {volume} {53}},\ \bibinfo {pages} {811} (\bibinfo {year}
  {1984})}\BibitemShut {NoStop}%
\bibitem [{\citenamefont {Stacy}\ and\ \citenamefont {Hodul}(1985)}]{Stacy85}%
  \BibitemOpen
  \bibfield  {author} {\bibinfo {author} {\bibfnamefont {A.~M.}\ \bibnamefont
  {Stacy}}\ and\ \bibinfo {author} {\bibfnamefont {D.~T.}\ \bibnamefont
  {Hodul}},\ }\href@noop {} {\bibfield  {journal} {\bibinfo  {journal} {J.
  Phys. Chem. Sol.}\ }\textbf {\bibinfo {volume} {46}},\ \bibinfo {pages} {405}
  (\bibinfo {year} {1985})}\BibitemShut {NoStop}%
\bibitem [{\citenamefont {Chen}\ and\ \citenamefont {Wang}(1974)}]{Chen74}%
  \BibitemOpen
  \bibfield  {author} {\bibinfo {author} {\bibfnamefont {J.~M.}\ \bibnamefont
  {Chen}}\ and\ \bibinfo {author} {\bibfnamefont {C.~S.}\ \bibnamefont
  {Wang}},\ }\href@noop {} {\bibfield  {journal} {\bibinfo  {journal} {Solid
  State Commun.}\ }\textbf {\bibinfo {volume} {14}},\ \bibinfo {pages} {857}
  (\bibinfo {year} {1974})}\BibitemShut {NoStop}%
\bibitem [{\citenamefont {Sourisseau}\ \emph {et~al.}(1989)\citenamefont
  {Sourisseau}, \citenamefont {Fouassier}, \citenamefont {Alba}, \citenamefont
  {Ghorayeb},\ and\ \citenamefont {Gorochov}}]{Sourisseau89}%
  \BibitemOpen
  \bibfield  {author} {\bibinfo {author} {\bibfnamefont {C.}~\bibnamefont
  {Sourisseau}}, \bibinfo {author} {\bibfnamefont {M.}~\bibnamefont
  {Fouassier}}, \bibinfo {author} {\bibfnamefont {M.}~\bibnamefont {Alba}},
  \bibinfo {author} {\bibfnamefont {A.}~\bibnamefont {Ghorayeb}},\ and\
  \bibinfo {author} {\bibfnamefont {O.}~\bibnamefont {Gorochov}},\ }\href@noop
  {} {\bibfield  {journal} {\bibinfo  {journal} {Mater. Sci. Eng.: B}\ }\textbf
  {\bibinfo {volume} {3}},\ \bibinfo {pages} {119} (\bibinfo {year}
  {1989})}\BibitemShut {NoStop}%
\bibitem [{\citenamefont {Feldman}\ \emph {et~al.}(1996)\citenamefont
  {Feldman}, \citenamefont {Frey}, \citenamefont {Homyonfer}, \citenamefont
  {Lyakhovitskaya}, \citenamefont {Margulis}, \citenamefont {Cohen},
  \citenamefont {Hodes}, \citenamefont {Hutchison},\ and\ \citenamefont
  {Tenne}}]{Feldman96}%
  \BibitemOpen
  \bibfield  {author} {\bibinfo {author} {\bibfnamefont {Y.}~\bibnamefont
  {Feldman}}, \bibinfo {author} {\bibfnamefont {G.~L.}\ \bibnamefont {Frey}},
  \bibinfo {author} {\bibfnamefont {M.}~\bibnamefont {Homyonfer}}, \bibinfo
  {author} {\bibfnamefont {V.}~\bibnamefont {Lyakhovitskaya}}, \bibinfo
  {author} {\bibfnamefont {L.}~\bibnamefont {Margulis}}, \bibinfo {author}
  {\bibfnamefont {H.}~\bibnamefont {Cohen}}, \bibinfo {author} {\bibfnamefont
  {G.}~\bibnamefont {Hodes}}, \bibinfo {author} {\bibfnamefont {J.~L.}\
  \bibnamefont {Hutchison}},\ and\ \bibinfo {author} {\bibfnamefont
  {R.}~\bibnamefont {Tenne}},\ }\href@noop {} {\bibfield  {journal} {\bibinfo
  {journal} {J. Am. Chem. Soc.}\ }\textbf {\bibinfo {volume} {118}},\ \bibinfo
  {pages} {5362} (\bibinfo {year} {1996})}\BibitemShut {NoStop}%
\bibitem [{\citenamefont {Frey}\ \emph {et~al.}(1999)\citenamefont {Frey},
  \citenamefont {Tenne}, \citenamefont {Matthews}, \citenamefont
  {Dresselhaus},\ and\ \citenamefont {Dresselhaus}}]{Frey99}%
  \BibitemOpen
  \bibfield  {author} {\bibinfo {author} {\bibfnamefont {G.~L.}\ \bibnamefont
  {Frey}}, \bibinfo {author} {\bibfnamefont {R.}~\bibnamefont {Tenne}},
  \bibinfo {author} {\bibfnamefont {M.~J.}\ \bibnamefont {Matthews}}, \bibinfo
  {author} {\bibfnamefont {M.~S.}\ \bibnamefont {Dresselhaus}},\ and\ \bibinfo
  {author} {\bibfnamefont {G.}~\bibnamefont {Dresselhaus}},\ }\href@noop {}
  {\bibfield  {journal} {\bibinfo  {journal} {Phys. Rev. B}\ }\textbf {\bibinfo
  {volume} {60}},\ \bibinfo {pages} {2883} (\bibinfo {year}
  {1999})}\BibitemShut {NoStop}%
\bibitem [{\citenamefont {Windom}\ \emph {et~al.}(2011)\citenamefont {Windom},
  \citenamefont {Sawyer},\ and\ \citenamefont {Hahn}}]{Windom11}%
  \BibitemOpen
  \bibfield  {author} {\bibinfo {author} {\bibfnamefont {B.~C.}\ \bibnamefont
  {Windom}}, \bibinfo {author} {\bibfnamefont {W.}~\bibnamefont {Sawyer}},\
  and\ \bibinfo {author} {\bibfnamefont {D.~W.}\ \bibnamefont {Hahn}},\
  }\href@noop {} {\bibfield  {journal} {\bibinfo  {journal} {Tribol. Lett.}\
  }\textbf {\bibinfo {volume} {42}},\ \bibinfo {pages} {301} (\bibinfo {year}
  {2011})}\BibitemShut {NoStop}%
\bibitem [{\citenamefont {Li}\ \emph {et~al.}(2012)\citenamefont {Li},
  \citenamefont {Zhang}, \citenamefont {Yap}, \citenamefont {Tay},
  \citenamefont {Edwin}, \citenamefont {Olivier},\ and\ \citenamefont
  {Baillargeat}}]{Li12}%
  \BibitemOpen
  \bibfield  {author} {\bibinfo {author} {\bibfnamefont {H.}~\bibnamefont
  {Li}}, \bibinfo {author} {\bibfnamefont {Q.}~\bibnamefont {Zhang}}, \bibinfo
  {author} {\bibfnamefont {C.~C.~R.}\ \bibnamefont {Yap}}, \bibinfo {author}
  {\bibfnamefont {B.~K.}\ \bibnamefont {Tay}}, \bibinfo {author} {\bibfnamefont
  {T.~H.~T.}\ \bibnamefont {Edwin}}, \bibinfo {author} {\bibfnamefont
  {A.}~\bibnamefont {Olivier}},\ and\ \bibinfo {author} {\bibfnamefont
  {D.}~\bibnamefont {Baillargeat}},\ }\href@noop {} {\bibfield  {journal}
  {\bibinfo  {journal} {Adv. Funct. Mater.}\ }\textbf {\bibinfo {volume}
  {22}},\ \bibinfo {pages} {1385} (\bibinfo {year} {2012})}\BibitemShut
  {NoStop}%
\bibitem [{\citenamefont {Chakraborty}\ \emph {et~al.}(2013)\citenamefont
  {Chakraborty}, \citenamefont {Matte}, \citenamefont {Sood},\ and\
  \citenamefont {Rao}}]{Chakraborty13}%
  \BibitemOpen
  \bibfield  {author} {\bibinfo {author} {\bibfnamefont {B.}~\bibnamefont
  {Chakraborty}}, \bibinfo {author} {\bibfnamefont {H.~S. S.~R.}\ \bibnamefont
  {Matte}}, \bibinfo {author} {\bibfnamefont {A.~K.}\ \bibnamefont {Sood}},\
  and\ \bibinfo {author} {\bibfnamefont {C.~N.~R.}\ \bibnamefont {Rao}},\
  }\href@noop {} {\bibfield  {journal} {\bibinfo  {journal} {J. Raman
  Spectrosc.}\ }\textbf {\bibinfo {volume} {44}},\ \bibinfo {pages} {92}
  (\bibinfo {year} {2013})}\BibitemShut {NoStop}%
\bibitem [{\citenamefont {Terrones}\ \emph {et~al.}(2014)\citenamefont
  {Terrones}, \citenamefont {Del~Corro}, \citenamefont {Feng}, \citenamefont
  {Poumirol}, \citenamefont {Rhodes}, \citenamefont {Smirnov}, \citenamefont
  {Pradhan}, \citenamefont {Lin}, \citenamefont {Nguyen}, \citenamefont
  {El\'{\i}as}, \citenamefont {Mallouk}, \citenamefont {Balicas}, \citenamefont
  {Pimenta},\ and\ \citenamefont {Terrones}}]{Terrones14}%
  \BibitemOpen
  \bibfield  {author} {\bibinfo {author} {\bibfnamefont {H.}~\bibnamefont
  {Terrones}}, \bibinfo {author} {\bibfnamefont {E.}~\bibnamefont {Del~Corro}},
  \bibinfo {author} {\bibfnamefont {S.}~\bibnamefont {Feng}}, \bibinfo {author}
  {\bibfnamefont {J.~M.}\ \bibnamefont {Poumirol}}, \bibinfo {author}
  {\bibfnamefont {D.}~\bibnamefont {Rhodes}}, \bibinfo {author} {\bibfnamefont
  {D.}~\bibnamefont {Smirnov}}, \bibinfo {author} {\bibfnamefont {N.~R.}\
  \bibnamefont {Pradhan}}, \bibinfo {author} {\bibfnamefont {Z.}~\bibnamefont
  {Lin}}, \bibinfo {author} {\bibfnamefont {M.~A.}\ \bibnamefont {Nguyen}},
  \bibinfo {author} {\bibfnamefont {A.~L.}\ \bibnamefont {El\'{\i}as}},
  \bibinfo {author} {\bibfnamefont {T.~E.}\ \bibnamefont {Mallouk}}, \bibinfo
  {author} {\bibfnamefont {L.}~\bibnamefont {Balicas}}, \bibinfo {author}
  {\bibfnamefont {M.~A.}\ \bibnamefont {Pimenta}},\ and\ \bibinfo {author}
  {\bibfnamefont {M.}~\bibnamefont {Terrones}},\ }\href@noop {} {\bibfield
  {journal} {\bibinfo  {journal} {Sci. Rep.}\ }\textbf {\bibinfo {volume}
  {4}},\ \bibinfo {pages} {4215} (\bibinfo {year} {2014})}\BibitemShut
  {NoStop}%
\bibitem [{\citenamefont {Berkdemir}\ \emph {et~al.}(2013)\citenamefont
  {Berkdemir}, \citenamefont {Guti$\acute{e}$rrez}, \citenamefont
  {Botello-M$\acute{e}$ndez}, \citenamefont {Perea-L$\acute{o}$pez},
  \citenamefont {El\'{\i}as}, \citenamefont {Chia}, \citenamefont {Wang},
  \citenamefont {Crespi}, \citenamefont {L$\acute{o}$pez-Ur\'{\i}as},
  \citenamefont {Charlier}, \citenamefont {Terrones},\ and\ \citenamefont
  {Terrones}}]{Berkdemir13}%
  \BibitemOpen
  \bibfield  {author} {\bibinfo {author} {\bibfnamefont {A.}~\bibnamefont
  {Berkdemir}}, \bibinfo {author} {\bibfnamefont {H.~R.}\ \bibnamefont
  {Guti$\acute{e}$rrez}}, \bibinfo {author} {\bibfnamefont {A.~R.}\
  \bibnamefont {Botello-M$\acute{e}$ndez}}, \bibinfo {author} {\bibfnamefont
  {N.}~\bibnamefont {Perea-L$\acute{o}$pez}}, \bibinfo {author} {\bibfnamefont
  {A.~L.}\ \bibnamefont {El\'{\i}as}}, \bibinfo {author} {\bibfnamefont
  {C.-I.}\ \bibnamefont {Chia}}, \bibinfo {author} {\bibfnamefont
  {B.}~\bibnamefont {Wang}}, \bibinfo {author} {\bibfnamefont {V.~H.}\
  \bibnamefont {Crespi}}, \bibinfo {author} {\bibfnamefont {F.}~\bibnamefont
  {L$\acute{o}$pez-Ur\'{\i}as}}, \bibinfo {author} {\bibfnamefont {J.-C.}\
  \bibnamefont {Charlier}}, \bibinfo {author} {\bibfnamefont {H.}~\bibnamefont
  {Terrones}},\ and\ \bibinfo {author} {\bibfnamefont {M.}~\bibnamefont
  {Terrones}},\ }\href@noop {} {\bibfield  {journal} {\bibinfo  {journal} {Sci.
  Rep.}\ }\textbf {\bibinfo {volume} {3}},\ \bibinfo {pages} {1755} (\bibinfo
  {year} {2013})}\BibitemShut {NoStop}%
\bibitem [{\citenamefont {Guo}\ \emph {et~al.}(2015)\citenamefont {Guo},
  \citenamefont {Yang}, \citenamefont {Yamamoto}, \citenamefont {Zhou},
  \citenamefont {Ishikawa}, \citenamefont {Ueno}, \citenamefont {Tsukagoshi},
  \citenamefont {Zhang}, \citenamefont {Dresselhaus},\ and\ \citenamefont
  {Saito}}]{Guo15}%
  \BibitemOpen
  \bibfield  {author} {\bibinfo {author} {\bibfnamefont {H.}~\bibnamefont
  {Guo}}, \bibinfo {author} {\bibfnamefont {T.}~\bibnamefont {Yang}}, \bibinfo
  {author} {\bibfnamefont {M.}~\bibnamefont {Yamamoto}}, \bibinfo {author}
  {\bibfnamefont {L.}~\bibnamefont {Zhou}}, \bibinfo {author} {\bibfnamefont
  {R.}~\bibnamefont {Ishikawa}}, \bibinfo {author} {\bibfnamefont
  {K.}~\bibnamefont {Ueno}}, \bibinfo {author} {\bibfnamefont {K.}~\bibnamefont
  {Tsukagoshi}}, \bibinfo {author} {\bibfnamefont {Z.}~\bibnamefont {Zhang}},
  \bibinfo {author} {\bibfnamefont {M.~S.}\ \bibnamefont {Dresselhaus}},\ and\
  \bibinfo {author} {\bibfnamefont {R.}~\bibnamefont {Saito}},\ }\href@noop {}
  {\bibfield  {journal} {\bibinfo  {journal} {Phys. Rev. B}\ }\textbf {\bibinfo
  {volume} {91}},\ \bibinfo {pages} {205415} (\bibinfo {year}
  {2015})}\BibitemShut {NoStop}%
\bibitem [{\citenamefont {Livneh}\ and\ \citenamefont
  {Spanier}(2015)}]{Livneh2015}%
  \BibitemOpen
  \bibfield  {author} {\bibinfo {author} {\bibfnamefont {T.}~\bibnamefont
  {Livneh}}\ and\ \bibinfo {author} {\bibfnamefont {J.~E.}\ \bibnamefont
  {Spanier}},\ }\href@noop {} {\bibfield  {journal} {\bibinfo  {journal} {2D
  Materials}\ }\textbf {\bibinfo {volume} {2}},\ \bibinfo {pages} {035003}
  (\bibinfo {year} {2015})}\BibitemShut {NoStop}%
\bibitem [{\citenamefont {Carvalho}\ \emph {et~al.}(2017)\citenamefont
  {Carvalho}, \citenamefont {Wang}, \citenamefont {Mignuzzi}, \citenamefont
  {Roy}, \citenamefont {Terrones}, \citenamefont {Fantini}, \citenamefont
  {Crespi}, \citenamefont {Malard},\ and\ \citenamefont
  {Pimenta}}]{Carvalho2017}%
  \BibitemOpen
  \bibfield  {author} {\bibinfo {author} {\bibfnamefont {B.}~\bibnamefont
  {Carvalho}}, \bibinfo {author} {\bibfnamefont {Y.}~\bibnamefont {Wang}},
  \bibinfo {author} {\bibfnamefont {S.}~\bibnamefont {Mignuzzi}}, \bibinfo
  {author} {\bibfnamefont {D.}~\bibnamefont {Roy}}, \bibinfo {author}
  {\bibfnamefont {M.}~\bibnamefont {Terrones}}, \bibinfo {author}
  {\bibfnamefont {C.}~\bibnamefont {Fantini}}, \bibinfo {author} {\bibfnamefont
  {V.~H.}\ \bibnamefont {Crespi}}, \bibinfo {author} {\bibfnamefont {L.~M.}\
  \bibnamefont {Malard}},\ and\ \bibinfo {author} {\bibfnamefont {M.~A.}\
  \bibnamefont {Pimenta}},\ }\href@noop {} {\bibfield  {journal} {\bibinfo
  {journal} {Nat. Commun.}\ }\textbf {\bibinfo {volume} {8}},\ \bibinfo {pages}
  {14670} (\bibinfo {year} {2017})}\BibitemShut {NoStop}%
\bibitem [{\citenamefont {Ferrari}\ and\ \citenamefont
  {Basko}(2013)}]{Ferrari2013}%
  \BibitemOpen
  \bibfield  {author} {\bibinfo {author} {\bibfnamefont {A.}~\bibnamefont
  {Ferrari}}\ and\ \bibinfo {author} {\bibfnamefont {D.}~\bibnamefont
  {Basko}},\ }\href@noop {} {\bibfield  {journal} {\bibinfo  {journal} {Nat.
  Nanotechnol.}\ }\textbf {\bibinfo {volume} {8}},\ \bibinfo {pages} {235}
  (\bibinfo {year} {2013})}\BibitemShut {NoStop}%
\bibitem [{\citenamefont {Liu}\ \emph {et~al.}(2015)\citenamefont {Liu},
  \citenamefont {Guo}, \citenamefont {Yang}, \citenamefont {Zhang},
  \citenamefont {Kumamoto}, \citenamefont {Shen}, \citenamefont {Hsu},
  \citenamefont {Li}, \citenamefont {Saito},\ and\ \citenamefont
  {Kawata}}]{Liu2015}%
  \BibitemOpen
  \bibfield  {author} {\bibinfo {author} {\bibfnamefont {H.-L.}\ \bibnamefont
  {Liu}}, \bibinfo {author} {\bibfnamefont {H.}~\bibnamefont {Guo}}, \bibinfo
  {author} {\bibfnamefont {T.}~\bibnamefont {Yang}}, \bibinfo {author}
  {\bibfnamefont {Z.}~\bibnamefont {Zhang}}, \bibinfo {author} {\bibfnamefont
  {Y.}~\bibnamefont {Kumamoto}}, \bibinfo {author} {\bibfnamefont {C.-C.}\
  \bibnamefont {Shen}}, \bibinfo {author} {\bibfnamefont {Y.-T.}\ \bibnamefont
  {Hsu}}, \bibinfo {author} {\bibfnamefont {L.-J.}\ \bibnamefont {Li}},
  \bibinfo {author} {\bibfnamefont {R.}~\bibnamefont {Saito}},\ and\ \bibinfo
  {author} {\bibfnamefont {S.}~\bibnamefont {Kawata}},\ }\href@noop {}
  {\bibfield  {journal} {\bibinfo  {journal} {Phys. Chem. Chem. Phys.}\
  }\textbf {\bibinfo {volume} {17}},\ \bibinfo {pages} {14561} (\bibinfo {year}
  {2015})}\BibitemShut {NoStop}%
\bibitem [{\citenamefont {Giannozzi}\ \emph {et~al.}(2017)\citenamefont
  {Giannozzi}, \citenamefont {Andreussi}, \citenamefont {Brumme}, \citenamefont
  {Bunau}, \citenamefont {Buongiorno~Nardelli}, \citenamefont {Calandra},
  \citenamefont {Car}, \citenamefont {Cavazzoni}, \citenamefont {Ceresoli},
  \citenamefont {Cococcioni}, \citenamefont {Colonna}, \citenamefont
  {Carnimeo}, \citenamefont {Dal~Corso}, \citenamefont {de~Gironcoli},
  \citenamefont {Delugas}, \citenamefont {DiStasio}, \citenamefont {Ferretti},
  \citenamefont {Floris}, \citenamefont {Fratesi}, \citenamefont {Fugallo},
  \citenamefont {Gebauer}, \citenamefont {Gerstmann}, \citenamefont {Giustino},
  \citenamefont {Gorni}, \citenamefont {Jia}, \citenamefont {Kawamura},
  \citenamefont {Ko}, \citenamefont {Kokalj}, \citenamefont
  {K\"{u}\c{c}\"{u}kbenli}, \citenamefont {Lazzeri}, \citenamefont {Marsili},
  \citenamefont {Marzari}, \citenamefont {Mauri}, \citenamefont {Nguyen},
  \citenamefont {Nguyen}, \citenamefont {Otero-de-la Roza}, \citenamefont
  {Paulatto}, \citenamefont {Ponc{\'{e}}}, \citenamefont {Rocca}, \citenamefont
  {Sabatini}, \citenamefont {Santra}, \citenamefont {Schlipf}, \citenamefont
  {Seitsonen}, \citenamefont {Smogunov}, \citenamefont {Timrov}, \citenamefont
  {Thonhauser}, \citenamefont {Umari}, \citenamefont {Vast}, \citenamefont
  {Wu},\ and\ \citenamefont {Baroni}}]{Giannozzi2017}%
  \BibitemOpen
  \bibfield  {author} {\bibinfo {author} {\bibfnamefont {P.}~\bibnamefont
  {Giannozzi}}, \bibinfo {author} {\bibfnamefont {O.}~\bibnamefont
  {Andreussi}}, \bibinfo {author} {\bibfnamefont {T.}~\bibnamefont {Brumme}},
  \bibinfo {author} {\bibfnamefont {O.}~\bibnamefont {Bunau}}, \bibinfo
  {author} {\bibfnamefont {M.}~\bibnamefont {Buongiorno~Nardelli}}, \bibinfo
  {author} {\bibfnamefont {M.}~\bibnamefont {Calandra}}, \bibinfo {author}
  {\bibfnamefont {R.}~\bibnamefont {Car}}, \bibinfo {author} {\bibfnamefont
  {C.}~\bibnamefont {Cavazzoni}}, \bibinfo {author} {\bibfnamefont
  {D.}~\bibnamefont {Ceresoli}}, \bibinfo {author} {\bibfnamefont
  {M.}~\bibnamefont {Cococcioni}}, \bibinfo {author} {\bibfnamefont
  {N.}~\bibnamefont {Colonna}}, \bibinfo {author} {\bibfnamefont
  {I.}~\bibnamefont {Carnimeo}}, \bibinfo {author} {\bibfnamefont
  {A.}~\bibnamefont {Dal~Corso}}, \bibinfo {author} {\bibfnamefont
  {S.}~\bibnamefont {de~Gironcoli}}, \bibinfo {author} {\bibfnamefont
  {P.}~\bibnamefont {Delugas}}, \bibinfo {author} {\bibfnamefont {R.~A.}\
  \bibnamefont {DiStasio}}, \bibinfo {author} {\bibfnamefont {A.}~\bibnamefont
  {Ferretti}}, \bibinfo {author} {\bibfnamefont {A.}~\bibnamefont {Floris}},
  \bibinfo {author} {\bibfnamefont {G.}~\bibnamefont {Fratesi}}, \bibinfo
  {author} {\bibfnamefont {G.}~\bibnamefont {Fugallo}}, \bibinfo {author}
  {\bibfnamefont {R.}~\bibnamefont {Gebauer}}, \bibinfo {author} {\bibfnamefont
  {U.}~\bibnamefont {Gerstmann}}, \bibinfo {author} {\bibfnamefont
  {F.}~\bibnamefont {Giustino}}, \bibinfo {author} {\bibfnamefont
  {T.}~\bibnamefont {Gorni}}, \bibinfo {author} {\bibfnamefont
  {J.}~\bibnamefont {Jia}}, \bibinfo {author} {\bibfnamefont {M.}~\bibnamefont
  {Kawamura}}, \bibinfo {author} {\bibfnamefont {H.-Y.}\ \bibnamefont {Ko}},
  \bibinfo {author} {\bibfnamefont {A.}~\bibnamefont {Kokalj}}, \bibinfo
  {author} {\bibfnamefont {E.}~\bibnamefont {K\"{u}\c{c}\"{u}kbenli}}, \bibinfo
  {author} {\bibfnamefont {M.}~\bibnamefont {Lazzeri}}, \bibinfo {author}
  {\bibfnamefont {M.}~\bibnamefont {Marsili}}, \bibinfo {author} {\bibfnamefont
  {N.}~\bibnamefont {Marzari}}, \bibinfo {author} {\bibfnamefont
  {F.}~\bibnamefont {Mauri}}, \bibinfo {author} {\bibfnamefont {N.~L.}\
  \bibnamefont {Nguyen}}, \bibinfo {author} {\bibfnamefont {H.-V.}\
  \bibnamefont {Nguyen}}, \bibinfo {author} {\bibfnamefont {A.}~\bibnamefont
  {Otero-de-la Roza}}, \bibinfo {author} {\bibfnamefont {L.}~\bibnamefont
  {Paulatto}}, \bibinfo {author} {\bibfnamefont {S.}~\bibnamefont
  {Ponc{\'{e}}}}, \bibinfo {author} {\bibfnamefont {D.}~\bibnamefont {Rocca}},
  \bibinfo {author} {\bibfnamefont {R.}~\bibnamefont {Sabatini}}, \bibinfo
  {author} {\bibfnamefont {B.}~\bibnamefont {Santra}}, \bibinfo {author}
  {\bibfnamefont {M.}~\bibnamefont {Schlipf}}, \bibinfo {author} {\bibfnamefont
  {A.~P.}\ \bibnamefont {Seitsonen}}, \bibinfo {author} {\bibfnamefont
  {A.}~\bibnamefont {Smogunov}}, \bibinfo {author} {\bibfnamefont
  {I.}~\bibnamefont {Timrov}}, \bibinfo {author} {\bibfnamefont
  {T.}~\bibnamefont {Thonhauser}}, \bibinfo {author} {\bibfnamefont
  {P.}~\bibnamefont {Umari}}, \bibinfo {author} {\bibfnamefont
  {N.}~\bibnamefont {Vast}}, \bibinfo {author} {\bibfnamefont {X.}~\bibnamefont
  {Wu}},\ and\ \bibinfo {author} {\bibfnamefont {S.}~\bibnamefont {Baroni}},\
  }\href@noop {} {\bibfield  {journal} {\bibinfo  {journal} {Journal of
  Physics: Condensed Matter}\ }\textbf {\bibinfo {volume} {29}},\ \bibinfo
  {pages} {465901} (\bibinfo {year} {2017})}\BibitemShut {NoStop}%
\bibitem [{\citenamefont {Herziger}\ \emph {et~al.}(2014)\citenamefont
  {Herziger}, \citenamefont {Calandra}, \citenamefont {Gava}, \citenamefont
  {May}, \citenamefont {Lazzeri}, \citenamefont {Mauri},\ and\ \citenamefont
  {Maultzsch}}]{Herziger2014}%
  \BibitemOpen
  \bibfield  {author} {\bibinfo {author} {\bibfnamefont {F.}~\bibnamefont
  {Herziger}}, \bibinfo {author} {\bibfnamefont {M.}~\bibnamefont {Calandra}},
  \bibinfo {author} {\bibfnamefont {P.}~\bibnamefont {Gava}}, \bibinfo {author}
  {\bibfnamefont {P.}~\bibnamefont {May}}, \bibinfo {author} {\bibfnamefont
  {M.}~\bibnamefont {Lazzeri}}, \bibinfo {author} {\bibfnamefont
  {F.}~\bibnamefont {Mauri}},\ and\ \bibinfo {author} {\bibfnamefont
  {J.}~\bibnamefont {Maultzsch}},\ }\href
  {https://doi.org/10.1103/PhysRevLett.113.187401} {\bibfield  {journal}
  {\bibinfo  {journal} {Phys. Rev. Lett.}\ }\textbf {\bibinfo {volume} {113}},\
  \bibinfo {pages} {187401} (\bibinfo {year} {2014})}\BibitemShut {NoStop}%
\bibitem [{\citenamefont {Torche}\ \emph {et~al.}(2017)\citenamefont {Torche},
  \citenamefont {Mauri}, \citenamefont {Charlier},\ and\ \citenamefont
  {Calandra}}]{Torche2017}%
  \BibitemOpen
  \bibfield  {author} {\bibinfo {author} {\bibfnamefont {A.}~\bibnamefont
  {Torche}}, \bibinfo {author} {\bibfnamefont {F.}~\bibnamefont {Mauri}},
  \bibinfo {author} {\bibfnamefont {J.-C.}\ \bibnamefont {Charlier}},\ and\
  \bibinfo {author} {\bibfnamefont {M.}~\bibnamefont {Calandra}},\ }\href
  {https://doi.org/10.1103/PhysRevMaterials.1.041001} {\bibfield  {journal}
  {\bibinfo  {journal} {Phys. Rev. Materials}\ }\textbf {\bibinfo {volume}
  {1}},\ \bibinfo {pages} {041001(R)} (\bibinfo {year} {2017})}\BibitemShut
  {NoStop}%
\bibitem [{\citenamefont {Giannozzi}\ \emph {et~al.}(2009)\citenamefont
  {Giannozzi}, \citenamefont {Baroni}, \citenamefont {Bonini}, \citenamefont
  {Calandra}, \citenamefont {Car}, \citenamefont {Cavazzoni}, \citenamefont
  {Ceresoli}, \citenamefont {Chiarotti}, \citenamefont {Cococcioni},
  \citenamefont {Dabo}, \citenamefont {Corso}, \citenamefont {Gironcoli},
  \citenamefont {Fabris}, \citenamefont {Fratesi}, \citenamefont {Gebauer},
  \citenamefont {Gerstmann}, \citenamefont {Gougoussis}, \citenamefont
  {Kokalj}, \citenamefont {Lazzeri}, \citenamefont {Martin-Samos},
  \citenamefont {Marzari}, \citenamefont {Mauri}, \citenamefont {Mazzarello},
  \citenamefont {Paolini}, \citenamefont {Pasquarello}, \citenamefont
  {Paulatto}, \citenamefont {Sbraccia}, \citenamefont {Scandolo}, \citenamefont
  {Sclauzero}, \citenamefont {Seitsonen}, \citenamefont {Smogunov},
  \citenamefont {Umari},\ and\ \citenamefont {Wentzcovitch}}]{Giannozzi09}%
  \BibitemOpen
  \bibfield  {author} {\bibinfo {author} {\bibfnamefont {P.}~\bibnamefont
  {Giannozzi}}, \bibinfo {author} {\bibfnamefont {S.}~\bibnamefont {Baroni}},
  \bibinfo {author} {\bibfnamefont {N.}~\bibnamefont {Bonini}}, \bibinfo
  {author} {\bibfnamefont {M.}~\bibnamefont {Calandra}}, \bibinfo {author}
  {\bibfnamefont {R.}~\bibnamefont {Car}}, \bibinfo {author} {\bibfnamefont
  {C.}~\bibnamefont {Cavazzoni}}, \bibinfo {author} {\bibfnamefont
  {D.}~\bibnamefont {Ceresoli}}, \bibinfo {author} {\bibfnamefont {G.~L.}\
  \bibnamefont {Chiarotti}}, \bibinfo {author} {\bibfnamefont {M.}~\bibnamefont
  {Cococcioni}}, \bibinfo {author} {\bibfnamefont {I.}~\bibnamefont {Dabo}},
  \bibinfo {author} {\bibfnamefont {A.~D.}\ \bibnamefont {Corso}}, \bibinfo
  {author} {\bibfnamefont {S.~D.}\ \bibnamefont {Gironcoli}}, \bibinfo {author}
  {\bibfnamefont {S.}~\bibnamefont {Fabris}}, \bibinfo {author} {\bibfnamefont
  {G.}~\bibnamefont {Fratesi}}, \bibinfo {author} {\bibfnamefont
  {R.}~\bibnamefont {Gebauer}}, \bibinfo {author} {\bibfnamefont
  {U.}~\bibnamefont {Gerstmann}}, \bibinfo {author} {\bibfnamefont
  {C.}~\bibnamefont {Gougoussis}}, \bibinfo {author} {\bibfnamefont
  {A.}~\bibnamefont {Kokalj}}, \bibinfo {author} {\bibfnamefont
  {M.}~\bibnamefont {Lazzeri}}, \bibinfo {author} {\bibfnamefont
  {L.}~\bibnamefont {Martin-Samos}}, \bibinfo {author} {\bibfnamefont
  {N.}~\bibnamefont {Marzari}}, \bibinfo {author} {\bibfnamefont
  {F.}~\bibnamefont {Mauri}}, \bibinfo {author} {\bibfnamefont
  {R.}~\bibnamefont {Mazzarello}}, \bibinfo {author} {\bibfnamefont
  {S.}~\bibnamefont {Paolini}}, \bibinfo {author} {\bibfnamefont
  {A.}~\bibnamefont {Pasquarello}}, \bibinfo {author} {\bibfnamefont
  {L.}~\bibnamefont {Paulatto}}, \bibinfo {author} {\bibfnamefont
  {C.}~\bibnamefont {Sbraccia}}, \bibinfo {author} {\bibfnamefont
  {S.}~\bibnamefont {Scandolo}}, \bibinfo {author} {\bibfnamefont
  {G.}~\bibnamefont {Sclauzero}}, \bibinfo {author} {\bibfnamefont {A.~P.}\
  \bibnamefont {Seitsonen}}, \bibinfo {author} {\bibfnamefont {A.}~\bibnamefont
  {Smogunov}}, \bibinfo {author} {\bibfnamefont {P.}~\bibnamefont {Umari}},\
  and\ \bibinfo {author} {\bibfnamefont {R.~M.}\ \bibnamefont {Wentzcovitch}},\
  }\href {http://www.quantum-espresso.org} {\bibfield  {journal} {\bibinfo
  {journal} {J. Phys.: Condens. Matter}\ }\textbf {\bibinfo {volume} {21}},\
  \bibinfo {pages} {395502} (\bibinfo {year} {2009})}\BibitemShut {NoStop}%
\bibitem [{\citenamefont {Monkhorst}\ and\ \citenamefont
  {Pack}(1976)}]{Monkhorst76}%
  \BibitemOpen
  \bibfield  {author} {\bibinfo {author} {\bibfnamefont {H.~J.}\ \bibnamefont
  {Monkhorst}}\ and\ \bibinfo {author} {\bibfnamefont {J.~D.}\ \bibnamefont
  {Pack}},\ }\href@noop {} {\bibfield  {journal} {\bibinfo  {journal} {Phys.
  Rev. B}\ }\textbf {\bibinfo {volume} {13}},\ \bibinfo {pages} {5188}
  (\bibinfo {year} {1976})}\BibitemShut {NoStop}%
\bibitem [{\citenamefont {Baroni}\ \emph {et~al.}(2001)\citenamefont {Baroni},
  \citenamefont {de~Gironcoli}, \citenamefont {Dal~Corso},\ and\ \citenamefont
  {Giannozzi}}]{Baroni01}%
  \BibitemOpen
  \bibfield  {author} {\bibinfo {author} {\bibfnamefont {S.}~\bibnamefont
  {Baroni}}, \bibinfo {author} {\bibfnamefont {S.}~\bibnamefont
  {de~Gironcoli}}, \bibinfo {author} {\bibfnamefont {A.}~\bibnamefont
  {Dal~Corso}},\ and\ \bibinfo {author} {\bibfnamefont {P.}~\bibnamefont
  {Giannozzi}},\ }\href@noop {} {\bibfield  {journal} {\bibinfo  {journal}
  {Rev. Mod. Phys.}\ }\textbf {\bibinfo {volume} {73}},\ \bibinfo {pages} {515}
  (\bibinfo {year} {2001})}\BibitemShut {NoStop}%
\bibitem [{\citenamefont {Noffsinger}\ \emph {et~al.}(2010)\citenamefont
  {Noffsinger}, \citenamefont {Giustino}, \citenamefont {Malone}, \citenamefont
  {Park}, \citenamefont {Louie},\ and\ \citenamefont {Cohen}}]{Noffsinger2010}%
  \BibitemOpen
  \bibfield  {author} {\bibinfo {author} {\bibfnamefont {J.}~\bibnamefont
  {Noffsinger}}, \bibinfo {author} {\bibfnamefont {F.}~\bibnamefont
  {Giustino}}, \bibinfo {author} {\bibfnamefont {B.~D.}\ \bibnamefont
  {Malone}}, \bibinfo {author} {\bibfnamefont {C.-H.}\ \bibnamefont {Park}},
  \bibinfo {author} {\bibfnamefont {S.~G.}\ \bibnamefont {Louie}},\ and\
  \bibinfo {author} {\bibfnamefont {M.~L.}\ \bibnamefont {Cohen}},\ }\href@noop
  {} {\bibfield  {journal} {\bibinfo  {journal} {Computer Physics
  Communications}\ }\textbf {\bibinfo {volume} {181}},\ \bibinfo {pages} {2140}
  (\bibinfo {year} {2010})}\BibitemShut {NoStop}%
\bibitem [{\citenamefont {Ponc$\acute{e}$}\ \emph {et~al.}(2016)\citenamefont
  {Ponc$\acute{e}$}, \citenamefont {Margine}, \citenamefont {Verdi},\ and\
  \citenamefont {Giustino}}]{Ponce2016}%
  \BibitemOpen
  \bibfield  {author} {\bibinfo {author} {\bibfnamefont {S.}~\bibnamefont
  {Ponc$\acute{e}$}}, \bibinfo {author} {\bibfnamefont {E.}~\bibnamefont
  {Margine}}, \bibinfo {author} {\bibfnamefont {C.}~\bibnamefont {Verdi}},\
  and\ \bibinfo {author} {\bibfnamefont {F.}~\bibnamefont {Giustino}},\
  }\href@noop {} {\bibfield  {journal} {\bibinfo  {journal} {Computer Physics
  Communications}\ }\textbf {\bibinfo {volume} {209}},\ \bibinfo {pages} {116}
  (\bibinfo {year} {2016})}\BibitemShut {NoStop}%
\bibitem [{\citenamefont {Lieth}(1977)}]{Lieth77}%
  \BibitemOpen
  \bibfield  {author} {\bibinfo {author} {\bibfnamefont {R.~M.~A.}\
  \bibnamefont {Lieth}},\ }\href@noop {} {\emph {\bibinfo {title} {Preparation
  and Crystal Growth of Materials with Layered Structures}}}\ (\bibinfo
  {publisher} {Springer: Berlin},\ \bibinfo {year} {1977})\BibitemShut
  {NoStop}%
\bibitem [{\citenamefont {Yamamoto}\ \emph {et~al.}(2014)\citenamefont
  {Yamamoto}, \citenamefont {Wang}, \citenamefont {Ni}, \citenamefont {Lin},
  \citenamefont {Li}, \citenamefont {Aikawa}, \citenamefont {Jian},
  \citenamefont {Ueno}, \citenamefont {Wakabayashi},\ and\ \citenamefont
  {Tsukagoshi}}]{yamamoto14}%
  \BibitemOpen
  \bibfield  {author} {\bibinfo {author} {\bibfnamefont {M.}~\bibnamefont
  {Yamamoto}}, \bibinfo {author} {\bibfnamefont {S.-T.}\ \bibnamefont {Wang}},
  \bibinfo {author} {\bibfnamefont {M.-Y.}\ \bibnamefont {Ni}}, \bibinfo
  {author} {\bibfnamefont {Y.-F.}\ \bibnamefont {Lin}}, \bibinfo {author}
  {\bibfnamefont {S.-L.}\ \bibnamefont {Li}}, \bibinfo {author} {\bibfnamefont
  {S.}~\bibnamefont {Aikawa}}, \bibinfo {author} {\bibfnamefont {W.-B.}\
  \bibnamefont {Jian}}, \bibinfo {author} {\bibfnamefont {K.}~\bibnamefont
  {Ueno}}, \bibinfo {author} {\bibfnamefont {K.}~\bibnamefont {Wakabayashi}},\
  and\ \bibinfo {author} {\bibfnamefont {K.}~\bibnamefont {Tsukagoshi}},\
  }\href@noop {} {\bibfield  {journal} {\bibinfo  {journal} {ACS Nano}\
  }\textbf {\bibinfo {volume} {8}},\ \bibinfo {pages} {3895} (\bibinfo {year}
  {2014})}\BibitemShut {NoStop}%
\bibitem [{\citenamefont {Spizzirri}(2010)}]{Spizzirri2010}%
  \BibitemOpen
  \bibfield  {author} {\bibinfo {author} {\bibfnamefont {P.~G.}\ \bibnamefont
  {Spizzirri}},\ }in\ \href@noop {} {\emph {\bibinfo {booktitle} {Microscopy:
  Science, Technology, Applications and Education}}},\ Vol.~\bibinfo {volume}
  {2},\ \bibinfo {editor} {edited by\ \bibinfo {editor} {\bibfnamefont
  {A.}~\bibnamefont {M{\'e}ndez-Vilas}}\ and\ \bibinfo {editor} {\bibfnamefont
  {J.}~\bibnamefont {D{\'\i}az}}}\ (\bibinfo  {publisher} {Formatex Research
  Center},\ \bibinfo {year} {2010})\ pp.\ \bibinfo {pages}
  {1389--1396}\BibitemShut {NoStop}%
\bibitem [{\citenamefont {Saito}\ \emph {et~al.}(2001)\citenamefont {Saito},
  \citenamefont {Jorio}, \citenamefont {Souza~Filho}, \citenamefont
  {Dresselhaus}, \citenamefont {Dresselhaus},\ and\ \citenamefont
  {Pimenta}}]{Saito02}%
  \BibitemOpen
  \bibfield  {author} {\bibinfo {author} {\bibfnamefont {R.}~\bibnamefont
  {Saito}}, \bibinfo {author} {\bibfnamefont {A.}~\bibnamefont {Jorio}},
  \bibinfo {author} {\bibfnamefont {A.~G.}\ \bibnamefont {Souza~Filho}},
  \bibinfo {author} {\bibfnamefont {G.}~\bibnamefont {Dresselhaus}}, \bibinfo
  {author} {\bibfnamefont {M.~S.}\ \bibnamefont {Dresselhaus}},\ and\ \bibinfo
  {author} {\bibfnamefont {M.~A.}\ \bibnamefont {Pimenta}},\ }\href@noop {}
  {\bibfield  {journal} {\bibinfo  {journal} {Phys. Rev. Lett.}\ }\textbf
  {\bibinfo {volume} {88}},\ \bibinfo {pages} {027401} (\bibinfo {year}
  {2001})}\BibitemShut {NoStop}%
\bibitem [{\citenamefont {Venezuela}\ \emph {et~al.}(2011)\citenamefont
  {Venezuela}, \citenamefont {Lazzeri},\ and\ \citenamefont
  {Mauri}}]{Venezuela11}%
  \BibitemOpen
  \bibfield  {author} {\bibinfo {author} {\bibfnamefont {P.}~\bibnamefont
  {Venezuela}}, \bibinfo {author} {\bibfnamefont {M.}~\bibnamefont {Lazzeri}},\
  and\ \bibinfo {author} {\bibfnamefont {F.}~\bibnamefont {Mauri}},\
  }\href@noop {} {\bibfield  {journal} {\bibinfo  {journal} {Phys. Rev. B}\
  }\textbf {\bibinfo {volume} {84}},\ \bibinfo {pages} {035433} (\bibinfo
  {year} {2011})}\BibitemShut {NoStop}%
\bibitem [{\citenamefont {Birman}(1962)}]{Birman1962}%
  \BibitemOpen
  \bibfield  {author} {\bibinfo {author} {\bibfnamefont {J.~L.}\ \bibnamefont
  {Birman}},\ }\href@noop {} {\bibfield  {journal} {\bibinfo  {journal} {Phys.
  Rev.}\ }\textbf {\bibinfo {volume} {127}},\ \bibinfo {pages} {1093} (\bibinfo
  {year} {1962})}\BibitemShut {NoStop}%
\bibitem [{\citenamefont {Birman}(1963)}]{Birman1963}%
  \BibitemOpen
  \bibfield  {author} {\bibinfo {author} {\bibfnamefont {J.~L.}\ \bibnamefont
  {Birman}},\ }\href@noop {} {\bibfield  {journal} {\bibinfo  {journal} {Phys.
  Rev.}\ }\textbf {\bibinfo {volume} {131}},\ \bibinfo {pages} {1489} (\bibinfo
  {year} {1963})}\BibitemShut {NoStop}%
\bibitem [{\citenamefont {Aroyo}\ \emph {et~al.}(2006)\citenamefont {Aroyo},
  \citenamefont {Kirov}, \citenamefont {Capillas}, \citenamefont {Perez-Mato},\
  and\ \citenamefont {Wondratschek}}]{Aroyo2006}%
  \BibitemOpen
  \bibfield  {author} {\bibinfo {author} {\bibfnamefont {M.~I.}\ \bibnamefont
  {Aroyo}}, \bibinfo {author} {\bibfnamefont {A.}~\bibnamefont {Kirov}},
  \bibinfo {author} {\bibfnamefont {C.}~\bibnamefont {Capillas}}, \bibinfo
  {author} {\bibfnamefont {J.~M.}\ \bibnamefont {Perez-Mato}},\ and\ \bibinfo
  {author} {\bibfnamefont {H.}~\bibnamefont {Wondratschek}},\ }\href@noop {}
  {\bibfield  {journal} {\bibinfo  {journal} {Acta Cryst. A}\ }\textbf
  {\bibinfo {volume} {62}},\ \bibinfo {pages} {115} (\bibinfo {year}
  {2006})}\BibitemShut {NoStop}%
\bibitem [{\citenamefont {Caramazza}\ \emph {et~al.}(2018)\citenamefont
  {Caramazza}, \citenamefont {Collina}, \citenamefont {Stellino}, \citenamefont
  {Ripanti}, \citenamefont {Dore},\ and\ \citenamefont
  {Postorino}}]{Caramazza2018}%
  \BibitemOpen
  \bibfield  {author} {\bibinfo {author} {\bibfnamefont {S.}~\bibnamefont
  {Caramazza}}, \bibinfo {author} {\bibfnamefont {A.}~\bibnamefont {Collina}},
  \bibinfo {author} {\bibfnamefont {E.}~\bibnamefont {Stellino}}, \bibinfo
  {author} {\bibfnamefont {F.}~\bibnamefont {Ripanti}}, \bibinfo {author}
  {\bibfnamefont {P.}~\bibnamefont {Dore}},\ and\ \bibinfo {author}
  {\bibfnamefont {P.}~\bibnamefont {Postorino}},\ }\href@noop {} {\bibfield
  {journal} {\bibinfo  {journal} {Eur. Phys. J. B}\ }\textbf {\bibinfo {volume}
  {91}},\ \bibinfo {pages} {35} (\bibinfo {year} {2018})}\BibitemShut {NoStop}%
\end{thebibliography}
%

\end{document}